\documentclass[twocolumn]{aastex631}

\usepackage{xspace}
\usepackage{multirow}
\usepackage{natbib}
\usepackage{CJK}
\usepackage{float}
\usepackage{romannum}
\usepackage{enumitem}

\defcitealias{2020MNRAS.499.5862L}{L20}
\defcitealias{2019MNRAS.489.4233M}{M19}
\defcitealias{2009ApJ...702.1472S}{S09}
\defcitealias{2003ApJS..146..165S}{S03}
\newcommand\be{\begin{eqnarray}}
\newcommand\ee{\end{eqnarray}}
\newcommand{\Rmnum}[1]{\uppercase\expandafter{\romannumeral #1}}

\shorttitle{\ion{O}{6} in Milky Way-like galaxies}
\shortauthors{Zhang et al.}

\begin{document}
\begin{CJK*}{UTF8}{gbsn}

\title{Low- and High-velocity \ion{O}{6} in Milky Way-like Galaxies: the Role of Stellar Feedback}

\correspondingauthor{Xiaoxia Zhang; Hui Li; Taotao Fang}
\email{zhangxx@xmu.edu.cn; hliastro@tsinghua.edu.cn; fangt@xmu.edu.cn}

\author{Zhijie Zhang (张志杰)}
\affiliation{Department of Astronomy, Xiamen University, Xiamen, Fujian 361005, China}
\author[0000-0003-4832-9422]{Xiaoxia Zhang (张小霞)}
\affiliation{Department of Astronomy, Xiamen University, Xiamen, Fujian 361005, China}
\author[0000-0002-1253-2763]{Hui Li (李辉)}
\affiliation{Department of Astronomy, Tsinghua University, Beijing 100084, China}
\affiliation{Department of Astronomy, Columbia University, Manhattan, New York 10027, USA}
\author[0000-0002-2853-3808]{Taotao Fang (方陶陶)}
\affiliation{Department of Astronomy, Xiamen University, Xiamen, Fujian 361005, China}
\author[0000-0003-3230-3981]{Qingzheng Yu (余清正)}
\affiliation{Department of Astronomy, Xiamen University, Xiamen, Fujian 361005, China}
\author[0000-0002-2243-2790]{Yang Luo (罗阳)}
\affiliation{Department of Astronomy, Yunnan University, Kunming, Yunnan 650000, China}
\author[0000-0003-3816-7028]{Federico Marinacci}
\affiliation{Department of Physics and Astronomy "Augusto Righi", University of Bologna, via Gobetti 93/2, I-40129 Bologna, Italy}
\affiliation{INAF, Astrophysics and Space Science Observatory Bologna, Via P. Gobetti 93/3, I-40129 Bologna, Italy}
\author[0000-0002-3790-720X]{Laura V. Sales}
\affiliation{Department of Physics and Astronomy, University of California, Riverside, 900 University Avenue, Riverside, CA 92521, USA}
\author[0000-0002-5653-0786]{Paul Torrey}
\affiliation{Department of Astronomy, University of Virginia, 530 McCormick Road, Charlottesville, VA 22903, USA}
\author[0000-0001-8593-7692]{Mark Vogelsberger}
\affiliation{Department of Physics and Kavli Institute for Astrophysics and Space Research, Massachusetts Institute of Technology, Cambridge, MA 02139, USA}

\begin{abstract}

Milky Way-type galaxies are surrounded by a warm-hot gaseous halo containing a considerable amount of baryons and metals. The kinematics and spatial distribution of highly-ionized ion species such as \ion{O}{6} can be significantly affected by supernova (SN) explosions and early (pre-SN) stellar feedback (e.g., stellar winds, radiation pressure). 
Here, we investigate effects of stellar feedback on \ion{O}{6} absorptions in Milky Way-like galaxies by analyzing the suites of high-resolution hydrodynamical simulations under the framework of  {\it SMUGGLE}, a physically motivated subgrid interstellar medium and stellar feedback model for the moving-mesh code {\sc Arepo}. 
We find that the fiducial run with the full suite of stellar feedback and moderate star formation activities can reasonably reproduce Galactic \ion{O}{6} absorptions observed by space telescopes such as {\it FUSE}, including the scale height of low-velocity ($|v_{\rm LSR}|< 100\, \rm km~s^{-1}$) \ion{O}{6}, the column density $-$ line width relation for high-velocity ($100 \leq |v_{\rm LSR}|< 400\, \rm km~s^{-1}$) \ion{O}{6}, and the cumulative \ion{O}{6} column densities. 
In contrast, model variations with more intense star formation activities deviate from observations further. 
Additionally, we find that the run considering only SN feedback is in broad agreement with the observations, whereas in runs without SN feedback this agreement is absent, which indicates a dominant role of SN feedback in heating and accelerating interstellar \ion{O}{6}. This is consistent with the current picture that interstellar \ion{O}{6} is predominantly produced by collisional ionization where mechanical feedback can play a central role.  In contrast, photoionization is negligible for \ion{O}{6} production due to the lack of high-energy ($\gtrsim114\ {\rm eV}$) photons required. 

\end{abstract}


\section{Introduction} 

The multiphase gas within and surrounding galaxies including the Milky Way (MW) is an essential ingredient of galactic ecosystems that govern the galaxy evolution, and may contain a significant amount of baryons and metals in the form of the cold ($T\lesssim10^4\ \rm K$), warm ($T\sim10^5-10^6\ \rm K$), and hot ($T\gtrsim10^6\ \rm K$) gaseous phases \citep[e.g.,][and references therein]{2012ARA&A..50..491P, 2017ARA&A..55..389T}.  
The existence of a warm-hot Galactic corona was originally proposed by \citet{1956ApJ...124...20S} to provide pressure confinement to the neutral clouds that are $\sim1\ \rm kpc$ above the Galactic plane, and was later confirmed by detections of the soft X-ray background \citep[e.g.,][]{1968Natur.217...32B} and interstellar \ion{O}{6} absorptions \citep[e.g.,][]{1974ApJ...193L.121J, 1974ApJ...193L.127Y}. 
\citet{1956ApJ...124...20S} also pointed out that such diffuse gas can be studied via the resonance doublet absorption lines of lithium-like ions, e.g., \ion{O}{6}, \ion{N}{5}, and \ion{C}{4}. 
Plasmas in the temperature range of about $(1-5)\times10^5\ \rm K$ traced by these species can be produced via moderate shocks or rapid cooling of hotter coronal gas probed in X-rays. 
\ion{O}{6} $\lambda\lambda1032, 1038$ doublet is of special significance owing to the large oscillator strengths  \citep{2003ApJS..149..205M} and high cosmic abundance of oxygen. 
Under the condition of collisionally ionized equilibrium, \ion{O}{6} peaks in abundance at the temperature of $\sim 3\times10^5\ \rm K$ \citep{1993ApJS...88..253S}.

The first large-scale surveys of \ion{O}{6} absorption in the Milky Way were made by the {\it Far Ultraviolet Spectroscopic Explorer} \citep[{\it FUSE};][]{2000ApJ...538L...1M, 2000ApJ...538L...7S}.
{\it FUSE} detections of \ion{O}{6} absorption lines toward extragalactic objects (e.g., active galactic nuclei (AGNs)/quasars) and stars 
in the Galactic disk, Galactic halo, and Magellanic clouds, reveal a widespread but irregular distribution of interstellar \ion{O}{6} with a column density of $\log (N/\rm cm^{-2}) \sim 13.0-14.8$ \citep[e.g.,][]{2000ApJ...538L..27S, 2002ApJ...572..264H, 2003ApJS..146....1W, 2003ApJS..146..125S, 2003ApJS..146..165S, 2005ApJ...622..377O, 2008ApJS..176...59B, 2017MNRAS.464.4927S}. 
The \ion{O}{6} absorbers detected by {\it FUSE} and {\it Hubble Space Telescope} ({\it HST}) along the lines of sight (LOS) of quasars/stars move at various velocities with respect to the local standard of rest (LSR), i.e., $|v_{\rm LSR}|$ ranges from $< 100$ to  $\gtrsim 400\, \rm km~s^{-1}$ \citep[e.g.,][]{2000ApJ...538L..31S, 2000ApJ...538L..35M, 2003ApJS..146..165S, 2006ApJS..165..229F, 2007ApJ...657..271C, 2011ApJ...739..105S}.

Low-velocity (e.g., $|v_{\rm LSR}|< 100\, \rm km~s^{-1}$) \ion{O}{6} is believed to be an extension of the Galactic disk, inflated due to its relatively high temperature, and can be well approximated by an exponentially declined layer with a midplane density of $\lesssim 2\times10^8\, \rm cm^{-3}$ and a scale height of $\sim 2.3-4\, \rm kpc$ \citep[e.g.,][]{2000ApJ...538L..27S, 2003ApJS..146..125S, 2003ApJ...586.1019Z, 2004ApJ...607..309I, 2009ApJ...702.1472S}. 
In contrast, the nature of high-velocity (e.g., $100 \leq |v_{\rm LSR}|< 400\, \rm km~s^{-1}$) \ion{O}{6} as well as intermediate- and low-ions (e.g., \ion{O}{1}, \ion{C}{2}, \ion{Si}{2}, \ion{Mg}{2}, \ion{Si}{3}, \ion{C}{4}, \ion{Si}{4}) and atoms, the so called high-velocity clouds (HVCs), is still debated, largely owing to the highly uncertain distances for most cases. 
While some high-velocity \ion{O}{6} features are spatially and kinematically associated with known \ion{H}{1} structures (e.g., Complex C and the Magellanic Steam), some have no neutral counterparts detected \citep[e.g.,][]{2003ApJS..146..165S, 2003Natur.421..719N, 2004ApJ...602..738F, 2004ApJ...605..216C, 2005ApJ...623..196C}.  
In addition, the covering fraction of high-velocity \ion{O}{6} \citep[$\gtrsim 60\%$; e.g.,][]{2003ApJS..146..165S, 2006ApJS..165..229F} is found to be higher than that of neutral and moderately ionized HVCs \citep[$\sim 20\%-40\%$ for \ion{H}{1}, \ion{C}{4}, and \ion{Si}{4}; e.g.,][]{2002ApJ...580L..47L, 2013A&A...550A..87H}, indicating a spatially more extended distribution for highly ionized HVCs. 
Despite the multiple origins proposed for HVCs, for example, the Galactic fountain \citep[e.g.,][]{1976ApJ...205..762S, 1980ApJ...236..577B, 2006MNRAS.366..449F}, materials stripped or ejected from satellite galaxies \citep[e.g.,][]{2004ASSL..312..101P, 2013A&A...550A..87H}, and accretion from the intergalactic medium \citep[IGM; e.g.,][]{2015MNRAS.447L..70F, 2009ApJ...700L...1K},  the spatial distribution and kinematics of high-velocity \ion{O}{6} are probably dominantly governed by the fountain model, which proposes that gas circulation in the halo is powered by stellar feedback, e.g., stellar winds and supernova (SN) explosions. Such a scenario is also supported by recently observed rain-like inflows and collimated outflows \citep[e.g.,][]{2022MNRAS.513.3228L, 2022MNRAS.515.4176M}.

\ion{O}{6} absorptions for low-redshift galaxies have been extensively studied by {\it HST} and {\it FUSE} \citep[e.g.,][]{2000ApJ...534L...1T, 2000ApJ...542...42T, 2005ApJ...624..555D, 2006ApJS..164....1L,  
2011Sci...334..948T, 2011ApJ...740...91P, 2011ApJ...743..180S, 2013ApJ...763..148S, 2013ApJ...778..187F, 2014ApJ...786...54P, 2014ApJ...792..128M, 2015MNRAS.449.3263J, 2015ApJ...815...22K, 2019ApJS..243...24P, 2022ApJ...927..147T}. 
Strong \ion{O}{6} absorptions have been preferentially detected around star-forming galaxies, with an average \ion{O}{6} column density of $\log (N/\rm cm^{-2}) \sim 14.5$ \citep[e.g.,][]{2011Sci...334..948T}, indicating a strong impact of star formation activities on the global properties of warm gaseous halo traced by \ion{O}{6} . Additionally, the covering fraction of \ion{O}{6} was found to depend on the inclination angle of galaxies and to follow a bimodal distribution that peaks within $\sim30^\circ$ of the galaxy minor axis and $\sim10^\circ-20^\circ$ of the major axis \citep[e.g.,][]{2015ApJ...815...22K}, consistent with a circumgalactic medium (CGM) originating from major-axis-fed inflows/recycled gas and from minor-axis-driven outflows, i.e., a scenario also revealed by cooler gaseous phases traced by \ion{Mg}{2} absorptions \citep[e.g.,][]{2012MNRAS.426..801B, 2012ApJ...760L...7K}. Those observational evidences highlight the influence of star formation activities and stellar/AGN feedback in shaping the spatial distribution of \ion{O}{6}-bearing gas of external galaxies. 

Stellar feedback, i.e., injection of substantial amounts of energy and angular momentum into the interstellar medium (ISM) via early (pre-SN) feedback and SN explosions, is likely to leave imprints on the properties of gaseous halos \citep[e.g.,][]{2021MNRAS.507.2383A, 2021A&A...655A..22M}. \ion{O}{6} ions probably trace matter in the interfaces between the cooler ionized/neutral clouds and hotter gas, and can thus serve as indirect probes of stellar feedback \citep[e.g.,][]{2011ApJ...727...46L}. 
Hydrodynamical simulations are powerful tools for studying the ISM/CGM, and given that multiscale physical processes are involved in galaxy formation, subgrid models are often used to implement small-scale processes such as star formation, metal mixing and transport, and stellar feedback \citep[e.g.,][]{1992ApJ...399L.113C, 2013ApJ...770...25A, 2014MNRAS.445..581H, 2017ApJ...834...69L, 2018ApJ...861..107L, 2018MNRAS.478..302S, 2018MNRAS.480..800H, 2019MNRAS.489.4233M}. Those subgruid models are parameterized and tuned to reproduce the observations, which means feedback energy is treated as a free parameter despite its well-known importance \citep[e.g.,][]{2023MNRAS.524.4091B}. 
A variety of simulations have shown that stellar feedback can have a significant impact on the physical properties such as kinematics, column densities, and total content of \ion{O}{6} \citep[e.g.,][]{2013MNRAS.430.1548H, 2015MNRAS.451.4223M, 2016MNRAS.458.1164L, 2017MNRAS.466.3810F, 2020ApJ...898..148L}.

Stars and MUltiphase Gas in GaLaxiEs \citep[{\it SMUGGLE};][]{2019MNRAS.489.4233M} is a physically motivated subgrid ISM and stellar feedback model for the moving-mesh code {\sc Arepo} \citep{2010MNRAS.401..791S} and has been widely used since its development \citep[e.g.,][]{2020MNRAS.499.5732K, 2022MNRAS.513.3458B, 2022MNRAS.517.4752S, 2023MNRAS.524.4091B}. 
It has successfully reproduced hydrogen emission line profile \citep[][]{2022MNRAS.517....1S}, constant density cores in dwarf galaxies \citep{2023MNRAS.520..461J}, and in particular, a realistic cold ISM and star cluster properties in isolated and merging galaxies \citep[][]{2020MNRAS.499.5862L, 2022MNRAS.514..265L}. 
In this paper, we test whether the {\it SMUGGLE} model can reproduce observations of warm \ion{O}{6} gas in and around the Milky Way, and investigate how the properties of \ion{O}{6} gas are affected by stellar feedback (e.g., early feedback and SN explosions) by analyzing the suites of simulations presented in \citet[][hereafter L20]{2020MNRAS.499.5862L}.

The paper is structured as follows. Section~\ref{sec:method} briefly introduces the {\it SMUGGLE} model and L20's simulation, and generates synthetic observations of \ion{O}{6} absorptions. Section~\ref{sec:result} presents results and discussion on \ion{O}{6} properties for different feedback model variations, as well as comparison with the observations and some caveats. Section~\ref{sec:summary} summarizes the main conclusions.

\begin{deluxetable*}{ccc}
\tablenum{1}
\tablecaption{Initial setup of the simulation performed in L20. \label{tab-1}}
\tablewidth{0pt}
\setlength{\tabcolsep}{10pt} 
\renewcommand{\arraystretch}{1.05} 
\tablehead{
\colhead{Parameter} & \colhead{Description} & \colhead{Value} 
}
\startdata
$M_{\rm total}$ 	& total mass in $\rm M_{\odot}$		& $1.6\times 10^{12}$   \\
$M_{\rm disk}$ 		& gaseous disk mass in $\rm M_{\odot}$      					& $9\times 10^{9}$   \\
$m_{\rm g}$ 		& the mass resolution of the gas cell in $\rm M_{\odot}$     	& $1.4\times 10^{3}$    \\
$L$ 				& simulated box size in kpc       						& 600   \\
$r_{\rm g}$ 		& scale length of gaseous disk in kpc      						& 6   \\
$\epsilon_{\rm g}$ 	& minimum gravitational softening length of gas cells in pc    & 3.6   \\
$n_{\rm th}$     	 	& density threshold for star formation in $\rm cm^{-3}$ 			& 100  \\  
\enddata
\end{deluxetable*}

\section{Methodology} 
\label{sec:method}

We analyze a suite of hydrodynamic simulations of isolated MW-sized galaxies presented in L20 under the {\it SMUGGLE} framework \citep{2019MNRAS.489.4233M}. We refer the reader to the original papers for details of the model and the simulations. Below we give a brief overview of the {\it SMUGGLE} model and L20's simulations, and describe the methodology we use to create mock observations of \ion{O}{6} absorptions toward background sources, following \citet{2002ApJ...564..604F}. The basic idea is to generate random LOS across the simulated region and obtain the temperature, baryon density, and velocity distributions along the LOS. Then \ion{O}{6} ion density can be derived from the metallicity and ionization fraction, from which the optical depth along the LOS and thus the synthetic spectrum can be obtained.  Finally, the column densities \ion{O}{6} and Doppler $b$-parameters for high- and low-velocity clouds can be calculated from the profile of \ion{O}{6} absorption line.

\subsection{The {\it SMUGGLE} galaxy formation model} 
 
The {\it SMUGGLE} model incorporates physical processes such as gravity, hydrodynamics, gas cooling and heating, star formation, and stellar feedback, and is able to resolve the multiphase gas structure of the ISM. Star particles are formed in cold, dense, and self-gravitating molecular gas reaching a density threshold of $n_{\rm th}=100\, \rm cm ^{-3}$. The local star formation rate for star-forming gas cells is controlled by the star formation efficiency per free-fall time $\epsilon_{\rm ff}$, i.e., $\dot{M}_*=\epsilon_{\rm ff} M_{\rm gas}/\tau_{\rm ff}$, with $M_{\rm gas}$ the gas mass and $\tau_{\rm ff}$ the free-fall time of the gas cell.

The model implements various channels of stellar feedback including photoionization, radiation pressure, energy and momentum injection from stellar winds and from supernovae, which are categorized into two types:

\begin{enumerate}[label=(\roman*), align=left]
\item {\it SN feedback} $-$ injects large amounts of energy and momentum to the ISM. The event number of type  \Rmnum{2} SNe at each time-step is obtained by integrating the \citet{2003PASP..115..763C} initial mass function, and the event rate of type \Rmnum{1}a SNe is calculated using a delay time distribution \citep{2013MNRAS.436.3031V}. 
\item {\it Early (pre-SN) feedback} $-$ includes radiative feedback and stellar winds. Photoionization and radiation pressure from young massive stars, namely radiative feedback, can impact the ionization state and offer pressure on surrounding gas and thus represent a source of momentum. 
The energy and momentum injection via stellar winds from young massive OB stars and older populations $-$ asymptotic giant branch (AGB) stars are calculated from the mass loss of the two types of stars, and the former provides another channel of early feedback.
\end{enumerate}

L20 performed a suite of high-resolution, isolated galactic disk simulations using the {\it SMUGGLE} model. Incorporated with explicit gas cooling and heating over a wide range of temperatures ($10 - 10^8\, \rm K$), the thermodynamical properties of the multiphase ISM is well studied. The simulations encompass a cubic region of $600\, \rm kpc$ on each side and cover the entire galaxy with the $z$-axis perpendicular to the galactic disk plane. The initial conditions of the simulation are the same as those of \citet{2019MNRAS.489.4233M}. It contains a  MW-sized galaxy of $1.6\times10^{12}\, \rm M_\odot$, which is composed of a stellar bulge and disk, a gaseous disk, and a dark matter halo, all with masses similar to those of the Milky Way \citep[see][and references therein]{2016ARA&A..54..529B}. The gaseous disk has an initial mass of $\sim 9\times10^9\, \rm M_\odot$ and the density decreases exponentially with a scale length of $6\, \rm kpc$. The initial setup leads to a gas fraction of roughly 10 percent within a radius of $R=8.5\, \rm kpc$. The mass resolution reaches $1.4\times10^3\, \rm M_\odot$ per gas cell, corresponding to the highest resolution run in \citet{2019MNRAS.489.4233M}. Gravitational softening is adaptive for gas cells, with a minimum value of $\sim3.6\ \rm pc$. Table~\ref{tab-1} lists the main parameters that characterize the initial condition of the simulations.

\begin{deluxetable}{cccc}
\tablenum{2}
\setlength{\tabcolsep}{13pt} 
\renewcommand{\arraystretch}{1.05} 
\tablecaption{Summarize of the six model variations in L20's simulations.  \label{tab-2}}
\tablewidth{0pt}
\tablehead{
\colhead{Run} & \colhead{$\epsilon_{\rm ff}$} &  \colhead{Radiation \& Winds} & \colhead{SN} 
}
\startdata
 SFE1 	& 0.01 	& Yes 	& Yes   \\
 SFE10 	& 0.1 	& Yes 	& Yes   \\
 SFE100 	& 1.0 	& Yes 	& Yes   \\
 Nofeed 	& 0.01 	& No 	& No   \\
 Rad 	& 0.01 	& Yes 	& No   \\  
 SN 		& 0.01 	& No 	& Yes   \\
 \enddata
\end{deluxetable}

In L20, we performed six simulations with different subgrid models (feedback channels) and parameters ($\epsilon_{\rm ff}$). The model variations are summarized in Table~\ref{tab-2} and detailed below. 

\begin{enumerate}[label=(\roman*), align=left]
	\item SFE1 $-$ fiducial run in \citetalias{2019MNRAS.489.4233M} with star formation efficiency of $\epsilon_{\rm ff}  = 0.01$ and all stellar feedback channels.
	\item SFE10 $-$ the same as SFE1 but with $\epsilon_{\rm ff}  = 0.1$.
	\item SFE100 $-$ the same as SFE1 but with $\epsilon_{\rm ff}  = 1$.
	\item Nofeed $-$ the same as SFE1 except with no stellar feedback.
	\item Rad $-$ the same as SFE1 except with only {\it early feedback} via stellar winds and radiation.
	\item SN $-$ the same as SFE1 except with only SN feedback.
\end{enumerate}

\subsection{Mock observations} 
\label{sec:obs}

To generate synthetic observational data, we build a mock galactic coordinate system consistent with that of the Milky Way. Specifically, we place the observer at the location of the Sun, i.e., $8.2\, \rm kpc$ away from the center of the simulated galaxy \citep{2016ARA&A..54..529B}. To avoid selection effects due to a single observer in a specific location, four observers at different off-center locations are situated on the galactic disk, each $8.2\, \rm kpc$ away from the galactic center and $90^\circ$ apart from each other \citep[similar to the eight off-center locations in][]{2020ApJ...896..143Z}. We define galactic longitude $l$ and latitude $b$ similar to those of the Galaxy. 

For each given set of ($l$, $b$) and distance $D$ of the star/quasar to the observer, we trace the LOS across the simulated region using the yt analysis toolkit \citep[http://yt-project.org;][]{2011ApJS..192....9T} that enables us to obtain gas properties such as temperatures, velocities in the LSR reference frame, and baryon densities along LOS. To directly compare the properties of the warm gas in the simulated galaxy with observations, we convert hydrogen density to \ion{O}{6} density, and then to \ion{O}{6} column density. For a grid of gas temperatures ($T\sim10^3 - 10^7\, \rm K$) and hydrogen densities ($n_{\rm H}\sim 10^{-8} - 10^6\, \rm cm^{-3}$), we adopt the {\sc Cloudy} code \citep[version C17.02;][]{1998PASP..110..761F, 2017RMxAA..53..385F} to calculate the ionization fraction $f_{\rm O \Romannum{6}} (T, n_{\rm H})$ of \ion{O}{6}, taking into account the ultraviolet (UV) background radiation from quasars and galaxies \citep{1996ApJ...461...20H}. The number density of \ion{O}{6} can be derived via 
\be 
n({\rm O\ \Romannum{6}})= n_{\rm H} A_{\rm O} \left(\frac{Z}{Z_\odot}\right)  f_{\rm O\ \Romannum{6}} (T, n_{\rm H}) ,
\label{eq-O6}
\ee
where $Z$ is the gas metallicity that is set to be the solar, i.e., $Z=Z_\odot$, and $A_{\rm O} = 4.9 \times 10^{-4}$ is the abundance of oxygen \citep{2009ARA&A..47..481A}. 

\begin{figure*}
\centering
\includegraphics[width=0.8\textwidth]{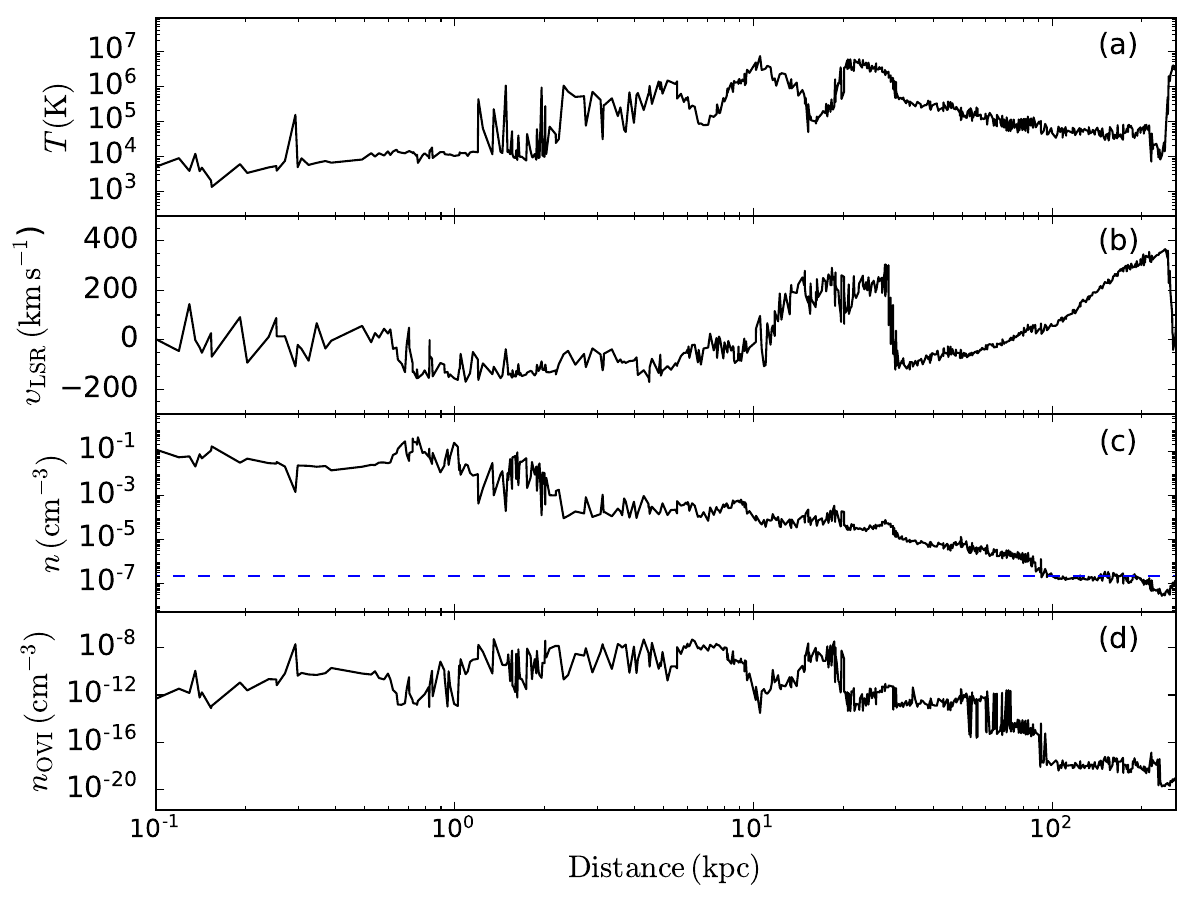}
\caption{Panels from top to bottom respectively represent the gas temperature (a), velocity in the LSR reference frame (b), baryon number density (c), and \ion{O}{6} number density (d) along the sightline $(l,\ b) = (0^\circ,\ 30^\circ)$ in the galactic coordinate to an observation at the Sun's location. The blue dashed line in panel (c) denotes the mean baryon density of the universe (see the text for details). 
\label{fig-Tvn}}
\end{figure*}

Take a random sightline at $(l,\ b) = (0^\circ,\ 30^\circ)$ as an example. Figure~\ref{fig-Tvn} shows the gas temperature, LSR velocity, baryon number density, and \ion{O}{6} number density along the sightline for the fiducial SFE1 run. The temperature of the gas along the LOS spans a wide range of $\sim10^3 - 10^7\, \rm K$, and the LSR velocity ranges from roughly $-200$ to $400\ \rm km\ s^{-1}$. The baryon density exhibits a downward trend as the distance increases, reaching the cosmic mean value of $\sim2.1\times 10^{-7}\,\rm cm^{-3}$ at $D\sim 100\, \rm kpc$ (the blue dashed line). \ion{O}{6} density generally traces the variation of gas temperature for distances $\lesssim 50\, \rm kpc$. 
The reason is that for temperature of $\lesssim5\times10^5\, \rm K$, the ionization fraction of \ion{O}{6} is a monotonic function of the gas temperature. 

\begin{figure}
\centering
\includegraphics[width=0.48\textwidth]{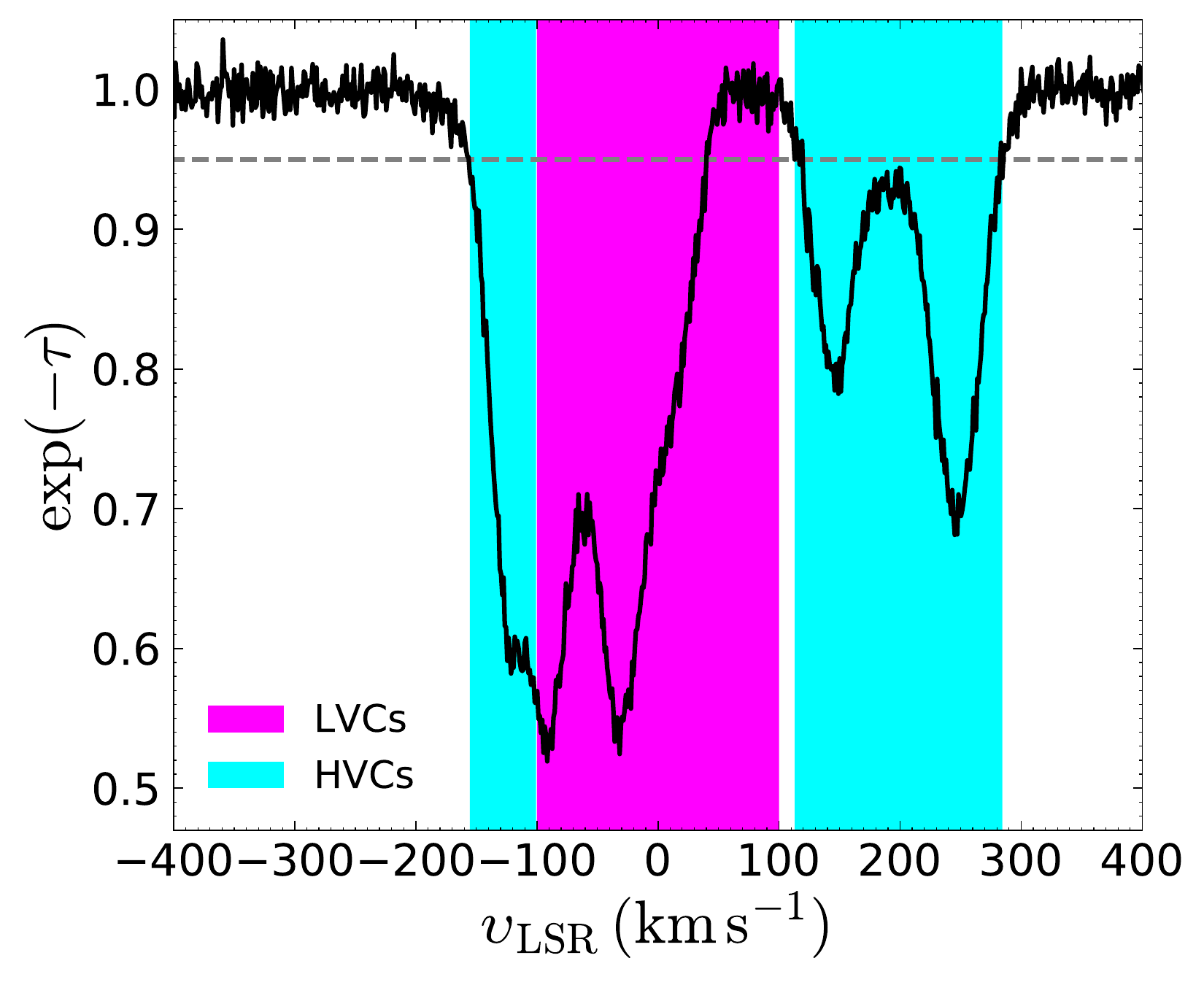}
\caption{An example of synthetic \ion{O}{6} absorption spectrum (the black solid line) and identification of low- and high-velocity \ion{O}{6} components. Velocities between $-100$ and $100\ \rm km~s^{-1}$ are identified as low-velocity component by the magenta band marks. High-velocity components are shown as the cyan bands, with velocities exceeding $100\ \rm km~s^{-1}$ and $\exp(-\tau)<0.95$. The gray dashed line marks $\exp(-\tau)=0.95$.   
\label{fig-spec}}
\end{figure}

The number density of \ion{O}{6} ions along the LOS provides a direct measure to the optical depth ($\tau$) around  \ion{O}{6} $1032\,\rm \AA$ line, from which the mock spectrum of a background source (e.g., a star or quasar) can be obtained \citep[e.g.,][]{1978ppim.book.....S,1997ApJ...485..496Z, 2002ApJ...564..604F}. We consider the effects of line broadening and line-center shift caused by both the Hubble velocity and the peculiar velocity of the gas along the LOS \citep[e.g.,][]{2002ApJ...564..604F}. 
The original spectrum is convolved with the {\it FUSE} line spread function to account for the instrumental broadening ($b_{\rm inst}$), i.e., $b = \sqrt{b_{\rm therm}^2 + b_{\rm inst}^2}$, with $b_{\rm therm}$ the thermal broadening and $b_{\rm inst} \sim 12-15\, \rm km~s ^{-1}$ \citep{2000ApJ...538L...1M, 2003ApJS..146..165S}. 
Gaussian noise is further considered with the mean of 0 and the standard deviation of 0.01. 
The final synthetic spectrum, or the transmission $\exp(-\tau)$, is shown as the black solid line in Figure~\ref{fig-spec}, for a background source placed at a distance of $260\, \rm kpc$ and in the direction of $(l,\ b) = (0^\circ,\ 30^\circ)$, the same LOS as in Figure~\ref{fig-Tvn}.
Multiple absorption components can be seen with $|v _{\rm LSR}| \sim 30-300\ \rm km~s^{-1}$, consistent with a wide spread of velocity for gas along the LOS shown in Figure~\ref{fig-Tvn}(b).

We adopt the apparent optical depth (AOD) method to calculate the column density ($N$), centroid velocity, and Doppler $b$-parameter for low- and high-velocity \ion{O}{6} by assuming that the absorption profile is not saturated \citep[e.g.,][]{1991ApJ...379..245S, 2003ApJS..146..165S}.  
For low-velocity \ion{O}{6}, the column density, centroid velocity, and line width are calculated with fixed integration limits of $(v_{-},\ v_{+}) = (-100,\ 100)\, \rm km~s^{-1}$ (the magenta band in Figure~\ref{fig-spec}). 
For high-velocity \ion{O}{6}, the integration limits rely on \ion{O}{6} velocity structures \citep[e.g.,][]{2003ApJS..146..165S} and are extracted from the cyan bands in Figure~\ref{fig-spec}, which include regions with $\exp(-\tau) < 0.95$ and $100 \le |v_{\rm LSR}| < 400\, \rm km~s^{-1}$.
Here $0.95$ is chosen somehow arbitrarily to exclude false ``absorption features'' caused by noise. The two cyan bands indicate two high-velocity components with integration limits of the velocity set by the boundaries of each cyan band.
We define $|v_{\rm LSR}| < 100\, \rm km~s^{-1}$ as low-velocity clouds (LVCs) and $100 \le |v_{\rm LSR}| < 400\, \rm km~s^{-1}$ as HVCs.

\begin{figure*}
\centering
\includegraphics[width=0.6\textwidth]{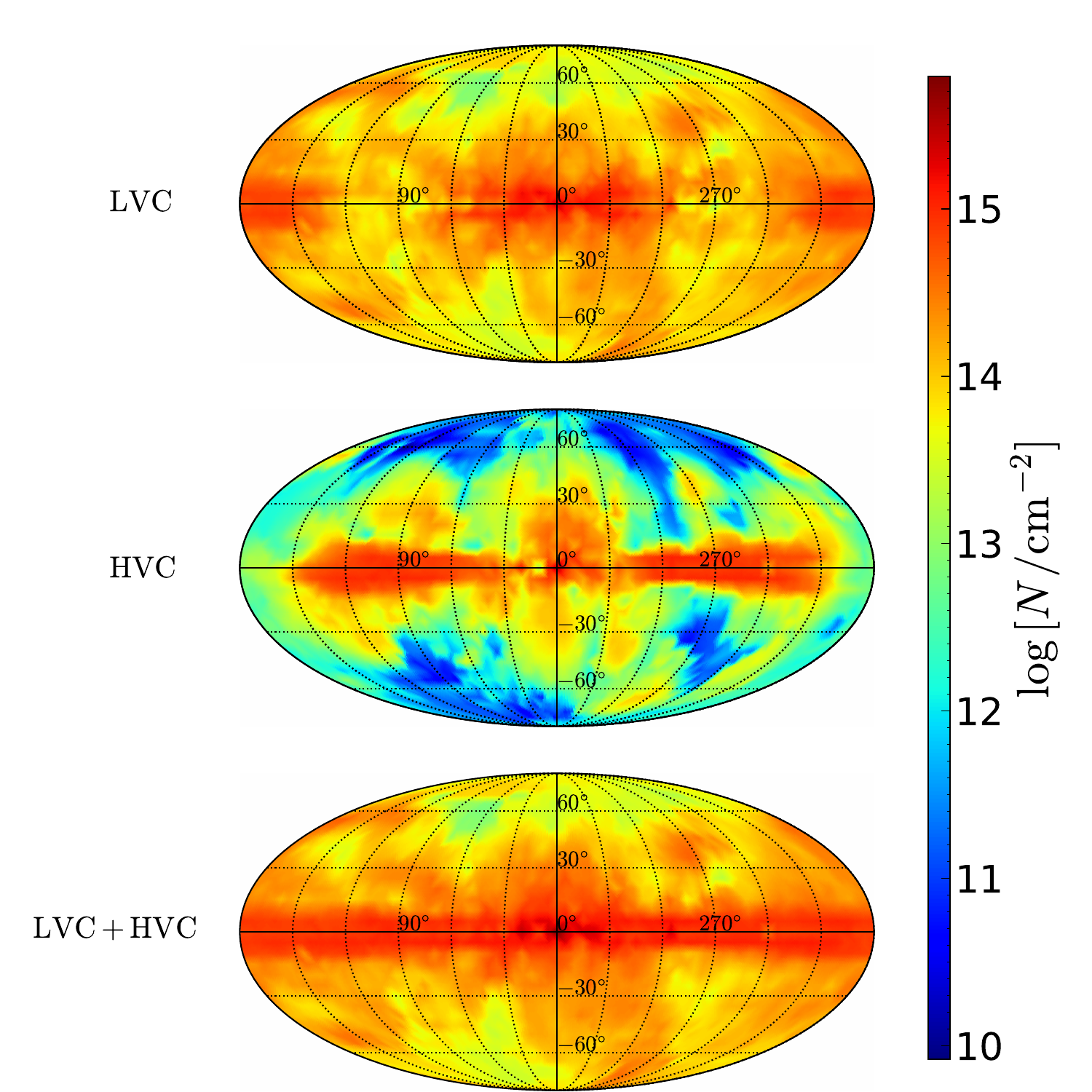}
\caption{All-sky Mollweide projection of \ion{O}{6} column densities for LVCs ($|v_{\rm LSR}| < 100\, \rm km~s^{-1}$; {\it top panel}), HVCs ($100 \le |v_{\rm LSR}| < 400\, \rm km~s^{-1}$; {\it middle panel}), and total clouds ($|v_{\rm LSR}| < 400\, \rm km~s^{-1}$; {\it bottom panel}), for an observer located at the Sun's position. 
\label{fig-map}}
\end{figure*}

Figure~\ref{fig-map} displays all-sky map of \ion{O}{6} column density derived from the fiducial SFE1 run,  for low-velocity (top), high-velocity (middle), and total gas (bottom), respectively, observed at the Sun's location. 
As can be seen, low-velocity \ion{O}{6} are widespread in the sky with column densities of $\log (N/\rm cm^{-2})\gtrsim14$, and stretch to high galactic latitude of $|b| \gtrsim 60^\circ$. 
In comparison, high-velocity \ion{O}{6} is generally located near the galactic disk with $\log (N/\rm cm^{-2})\gtrsim15$, and gradually decline toward higher galactic latitudes. 
Such a ``disk-like" structure for high-velocity \ion{O}{6} does not appear in the other five runs, which suggests that the spatial distribution of \ion{O}{6} strongly depends on the sub-grid models of stellar feedback. 
However, it is challenging to detect \ion{O}{6} absorption at low latitudes (e.g., $|b|\lesssim 25^\circ$) due to severe ultraviolet extinction for extragalactic objects \citep[e.g.,][]{2003ApJS..146....1W, 2003ApJS..146..165S}. Therefore, currently it is unavailable to distinguish those models via the observed spatial distribution of \ion{O}{6}.

\section{Results and discussion} 
\label{sec:result}

We derive \ion{O}{6} properties of the simulated galaxy viewed by four internal off-center observers for the fiducial SFE1 run in Section~\ref{sec:lvc} $-$ \ref{sec:cumulative}. The fiducial results are then compared with the other five model variations in Section~\ref{sec:other} where the impact of sub-grid model/parameter variations are illustrated. Section~\ref{sec:external} presents results from an external view to compare with observations of external galaxies, which is followed by some caveats in Section~\ref{sec:caveat}. 
  
\subsection{The scale height of LVCs} 
\label{sec:lvc}

The mock observation of the simulated galaxy in Section~\ref{sec:method} provides measurement to the \ion{O}{6} column density along arbitrary LOS across the simulated region. A certain number of sightlines allow us to explore the spatial distribution of the \ion{O}{6}-bearing gas, which can be compared with the observations. \citet{2009ApJ...702.1472S} collected column densities of \ion{O}{6} as well as other species along $139$ LOS toward stars and quasars, and found that low-velocity \ion{O}{6} in the Milky Way is well fitted by an exponentially declined disk model with a scale height of $h\sim2.6\pm0.6\, \rm kpc$. To compare our results with those observations, we generate random LOS according to the following settings.

\begin{figure}
\centering
\includegraphics[width=0.48\textwidth]{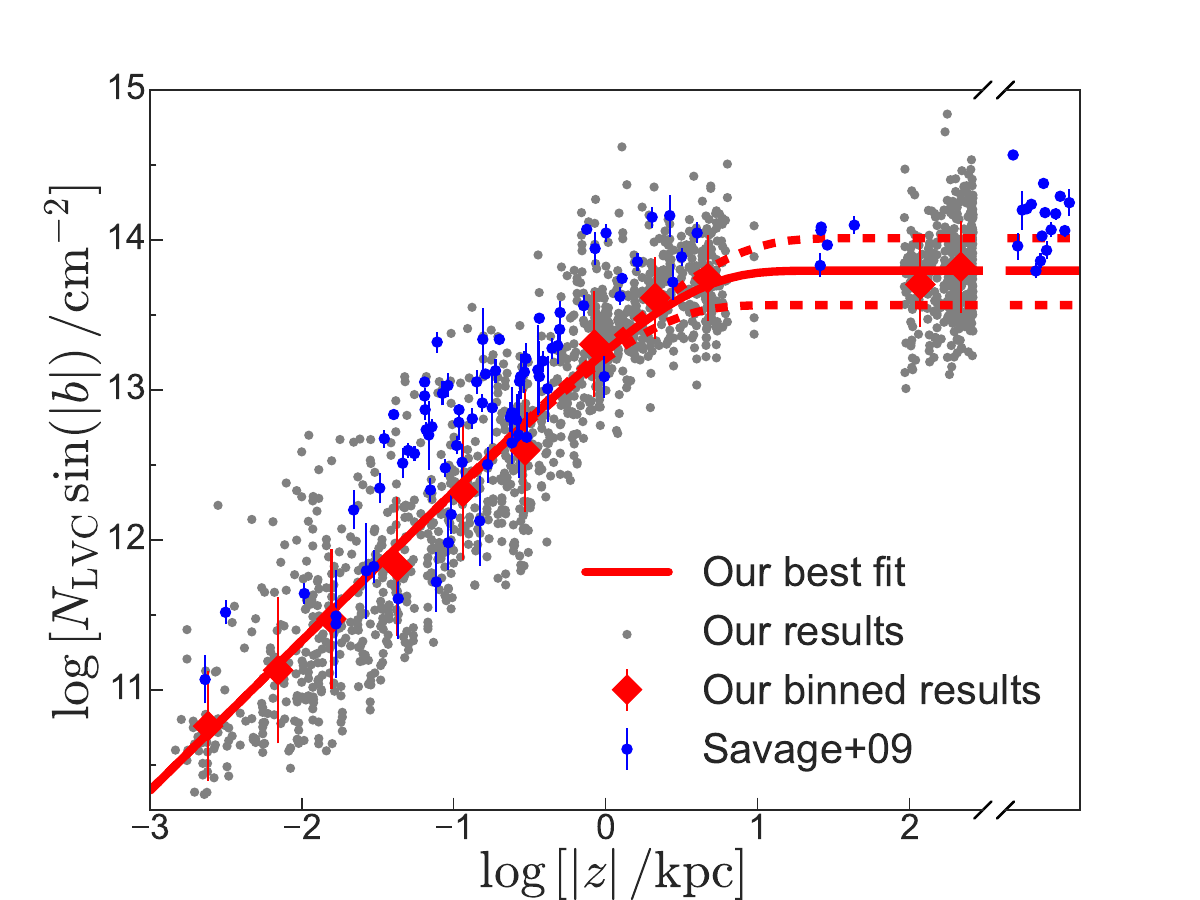}
\caption{Projected \ion{O}{6} column density along $z$-axis versus height to the galactic plane for LVCs in the SFE1 run. The gray dots represent our mock results, and the red diamonds are our binned results with $1\sigma$ error bars. The red solid line is the best-fit model to the red diamonds, and the dashed lines enclose $1\sigma$ confidence region accounting for the errors of the scale height. The blue circles are results revealed by observations \citep{2009ApJ...702.1472S}.
\label{fig-Nz}}
\end{figure}

For each of the four off-center observers, we randomly generate $4\times139$ LOS across the simulated galaxy toward quasars or stars, and thus the total sightline number is $4\times4\times139=2224$. Here $139$ is the LOS number collected by \citet{2009ApJ...702.1472S}, among which 109 (30) are toward stars (quasars). We assign the same ratio of numbers for sightlines toward quasars to that toward stars, i.e., 480 (1744) out of the $2224$ LOS are toward quasars (stars). 
The quasars are situated at a distance of $260\, \rm kpc$ (e.g., the virial radius of the Galaxy) and the galactic latitude is randomly drawn at $|b|>20^\circ$ since detectable sightlines toward quasars are usually observationally unavailable at $|b|\lesssim 20^\circ$.  The stars are placed in random directions with a distance randomly drawn from $1-10\, \rm kpc$ in  logarithmic space. This distance range is consistent with the observational data collected by \citet{2009ApJ...702.1472S}.
\ion{O}{6} column density for LVCs are derived according to the AOD method presented in Section~\ref{sec:obs}.  We further set a detection limit of \ion{O}{6} column density $\log (N/\rm cm^{-2}) \geq 13.23$ for sightlines toward both quasars and stars \citep[e.g.,][]{2009ApJ...702.1472S}, and there are $1601$ \ion{O}{6} absorbers detected along the $2224$ LOS, as the gray dots in Figure~\ref{fig-Nz} show. Our results generally agree with the observations of low-velocity \ion{O}{6} \citep[the blue symbols;][]{2009ApJ...702.1472S}.

To quantify the distribution of low-velocity \ion{O}{6}, we adopt a simple disk model \citep[e.g.,][]{1990ApJ...361..107S, 2000ApJ...538L..27S, 2005ApJ...624..751Y, 2009ApJ...702.1472S},  i.e., the number density declines exponentially away from the mid-plane (or the galactic disk), and the density at a height $z$ below/above the mid-plane can be expressed as 
\be
n(z)=n_0 e^{-|z|/h}, 
\label{eq-nz}
\ee
where $n_0$ is the mean density in the mid-plane, $h$ is the scale height of \ion{O}{6} disk. Then the column density ($N$) along the LOS can be simply derived from the integration of Equation~(\ref{eq-nz}), and its projection along $z$-axis is 
\be
N \sin |b| = n_0 h \left(1-e^{-|z|/h} \right).
\ee
The formula reveals a monotonic relation between $N \sin |b|$ and $|z|$, for $|z|\lesssim h$; and for $|z|\gg h$, $N \sin |b|$ eventually approaches a stable value of $n_0 h$.

To perform a reasonable minimum$-\chi^2$ fitting to our mock data (the gray dots in Figure~\ref{fig-Nz}), we divide the $x$-axis into bins. Red circles with error bars show the mean values and standard deviations in each bin. These binned data are then fitted to the disk model, with the best-fitting profile shown as the red solid line. Our best-fitting model has a scale height of $h = 2.9_{-1.2}^{+1.9}\, \rm kpc$, well consistent with the MW observations of $2.6\pm0.5\, \rm kpc$ \citep{2009ApJ...702.1472S} considering the $1\sigma$ errors. The best-fit parameters and a comparison with observations are listed in Table~\ref{tab-3}.

\begin{deluxetable*}{cccccccc}
\tablenum{3}
\tablecaption{Properties low- and high-velocity \ion{O}{6} clouds for the six runs in L20's simulation and comparison with the observations.} \label{tab-3}
\tablewidth{0pt}
\setlength{\tabcolsep}{8pt} 
\renewcommand{\arraystretch}{1.2} 
\tablehead{
\colhead{Model/Obs.} & \colhead{$h$} & \colhead{$n_0$} & \colhead{$\log(n_0 h)$} & \colhead{$b_{\rm LVC}$} & \colhead{$\log N_{\rm LVC}$} & \colhead{$b_{\rm HVC}$ } & \colhead{$\log N_{\rm HVC}$} \\
& \colhead{(kpc)} & \colhead{($\rm cm^{-3}$)} &\colhead{($\rm cm^{-2}$)}& \colhead{($\rm km\ s^{-1}$)} & \colhead{($\rm cm^{-2}$)}  & \colhead{($\rm km\ s^{-1}$)} &  \colhead{($\rm cm^{-2}$)}
}
\startdata
SFE1 	& $2.9^{+1.9}_{-1.2}$        & $6.92\times10^{-9}$       & $13.79 \pm 0.16$ 		& $47.38 \pm 15.18$ 	& $13.81 \pm 0.38$ 		  & $33.02 \pm 18.81$ 		& $13.80 \pm 0.37$ \\
SFE10 	& $1.3^{+1.3}_{-0.7}$        & $1.39\times10^{-8}$       & $ 13.77 \pm 0.18$  		& $47.62 \pm 17.29$ 	& $13.97 \pm 0.43$ 		& $20.17 \pm 9.56$ 		& $13.86 \pm 0.30$ \\
SFE100 	&  $1.9^{+1.6}_{-0.9}$       & $7.64\times10^{-9}$       & $ 13.67^{+0.16}_{-0.17}$ 		& $38.09 \pm 14.10$		& $13.79 \pm 0.41$ 		& $24.21 \pm 12.41$ 		& $13.66 \pm 0.29$ \\
Nofeed 	& $2.0^{+1.0}_{-0.7}$        & $4.10\times10^{-9}$       & $ 13.41^{+0.10}_{-0.11}$	  	& $39.56 \pm 18.81$ 	& $13.58 \pm 0.30$ 		&  	\nodata 			& \nodata \\
Rad 	& $0.5^{+1.6}_{-0.5}$          & $1.41\times10^{-8}$       & $ 13.36^{+0.14}_{-0.15}$  		& $39.59 \pm 16.27$ 	& $13.60 \pm 0.41$ 		& $27.28 \pm 5.31$ 		& $13.64 \pm 0.21$ \\  
SN 		& $2.7^{+2.2}_{-1.3}$       & $1.19\times10^{-8}$        & $ 13.99^{+0.20}_{-0.19}$  		& $47.89 \pm 17.90$ 	& $14.04 \pm 0.42$ 		& $31.95 \pm 20.11$ 		& $13.92 \pm 0.36$ \\
\hline
Savage+09 	& $2.6 \pm 0.5$ 		& $1.64\times10^{-8}$ 			& $14.12^{+0.07}_{-0.08}$   		& \nodata				& $14.15 \pm 0.35$ 		& \nodata				& \nodata \\  
Sembach+03 	& \nodata 				& \nodata 				& \nodata 				& \nodata 						&  \nodata				& $40.00 \pm 13.14$ 	& $13.97 \pm 0.33$ \\  
 \enddata
 \tablecomments{Columns from left to right represent: (1) the six model variations of the simulation presented in L20 (the second to seventh rows), or the observations of LVCs and HVCs by \citet{2009ApJ...702.1472S} and \citet{2003ApJS..146..165S}, respectively (the last two rows); (2)-(4) the best-fit parameters of the disk model for low-velocity \ion{O}{6}; (5)-(6) the median column density and line width for low-velocity \ion{O}{6}; (7)-(8) the median column density and line width for high-velocity \ion{O}{6}.}
\end{deluxetable*}

\subsection{Column density $-$ line width relation} 
\label{sec:hvc}

A correlation between the column density and the line width of  \ion{O}{6} absorbers was first reported by \citet{2002ApJ...577..691H} and has been found in various environments including the Galactic disk and halo \citep{1978ApJ...219..845J, 1978ApJ...220..107J, 2003ApJS..146..125S, 2008ApJS..176...59B, 2011ApJ...727...46L, 2017MNRAS.464.4927S}, HVCs \citep{2003ApJS..146..165S}, and Magellanic clouds \citep{2002ApJ...572..264H, 2002ApJ...569..214H, 2002ApJ...569..233H, 2011MNRAS.412.1105P}.  
Collisional processes should be responsible for the linear proportionality between the column density and $b$-parameter since the column density linearly scales with the gas flow velocity in collisional ionization scenario \citep[see discussions in][]{2002ApJ...577..691H, 2003ApJS..146..165S}. Below we investigate the column density $-$ line width relation for low- and high-velocity \ion{O}{6}, which are then compared with the observations.

\begin{figure}
\centering
\includegraphics[width=0.48\textwidth]{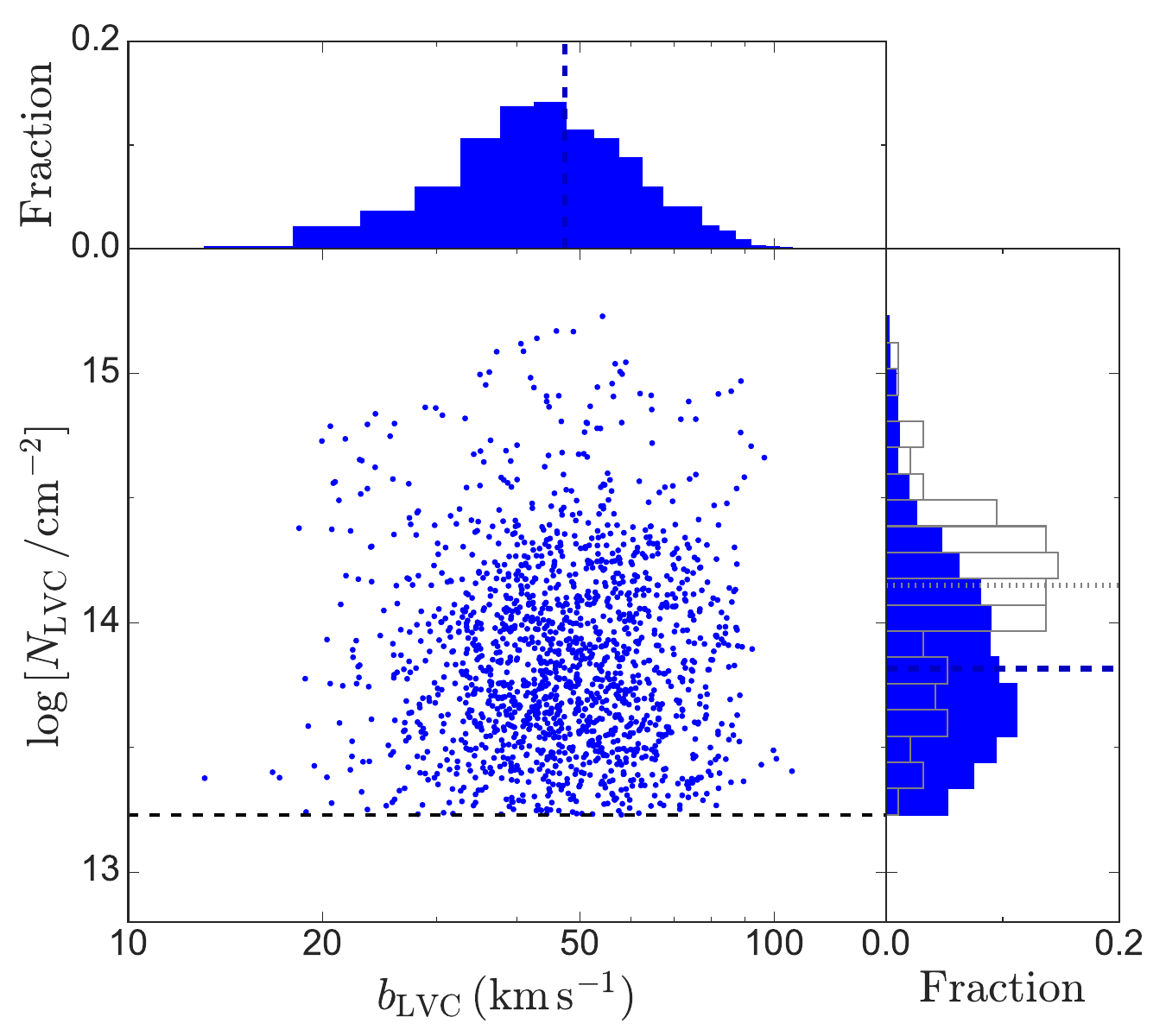}
\caption{Column density and line width distribution of \ion{O}{6}-bearing LVCs in the fiducial SFE1 run.  The blue histograms give the probability distribution of the column density (right) and line width (top), with the median values denoted by the blue dashed lines, as compared with the observations shown as the gray histogram and gray dotted line \citep{2009ApJ...702.1472S}. 
\label{fig-Nb-lo}}
\end{figure}

Figure~\ref{fig-Nb-lo} depicts the column density vs. Doppler parameter distribution for low-velocity \ion{O}{6}.  Each data point is obtained from a randomly drawn sightline at $|b|>20^\circ$, from four off-center observers toward quasars/stars, as described in Section~\ref{sec:lvc}. \ion{O}{6} column densities span $\log (N/\rm cm^{-2}) \sim 13.2- 15.2$ with a median value of $13.8$, and the line width follows a Gaussian-like distribution within $b\sim 13-106\, \rm km~s^{-1}$ and peaks at $\sim47\, \rm km~s^{-1}$. 
The median values are listed in the fourth and fifth columns in Table~\ref{tab-3}. The distribution and median value of the column density are well consistent with the observations \citep[the gray histogram and dotted line;][]{2009ApJ...702.1472S}. For most of the sightlines, the line width is broader than that caused by thermal motion of \ion{O}{6} ions, which corresponds to $b_{\rm therm} \sim 17.7\ \rm km~s^{-1}$ for gas temperature of $3\times10^5\ \rm K$,  implying significant non-thermal motions, e.g., inflows, outflows and turbulence.  
This could be responsible for the distorted or no relation between the column density and line width. Although no correlation between $N$ and $b$ is also expected for photoionized gas, given the high energy ($\sim114$ eV) required for ionizing photons, most of \ion{O}{6} ions are implausible to be produced by photoionization except for extreme conditions with a hard radiation field and a very low gas density \citep[see e.g.,][]{2003ApJS..146..165S}.

\begin{figure}
\centering
\includegraphics[width=0.48\textwidth]{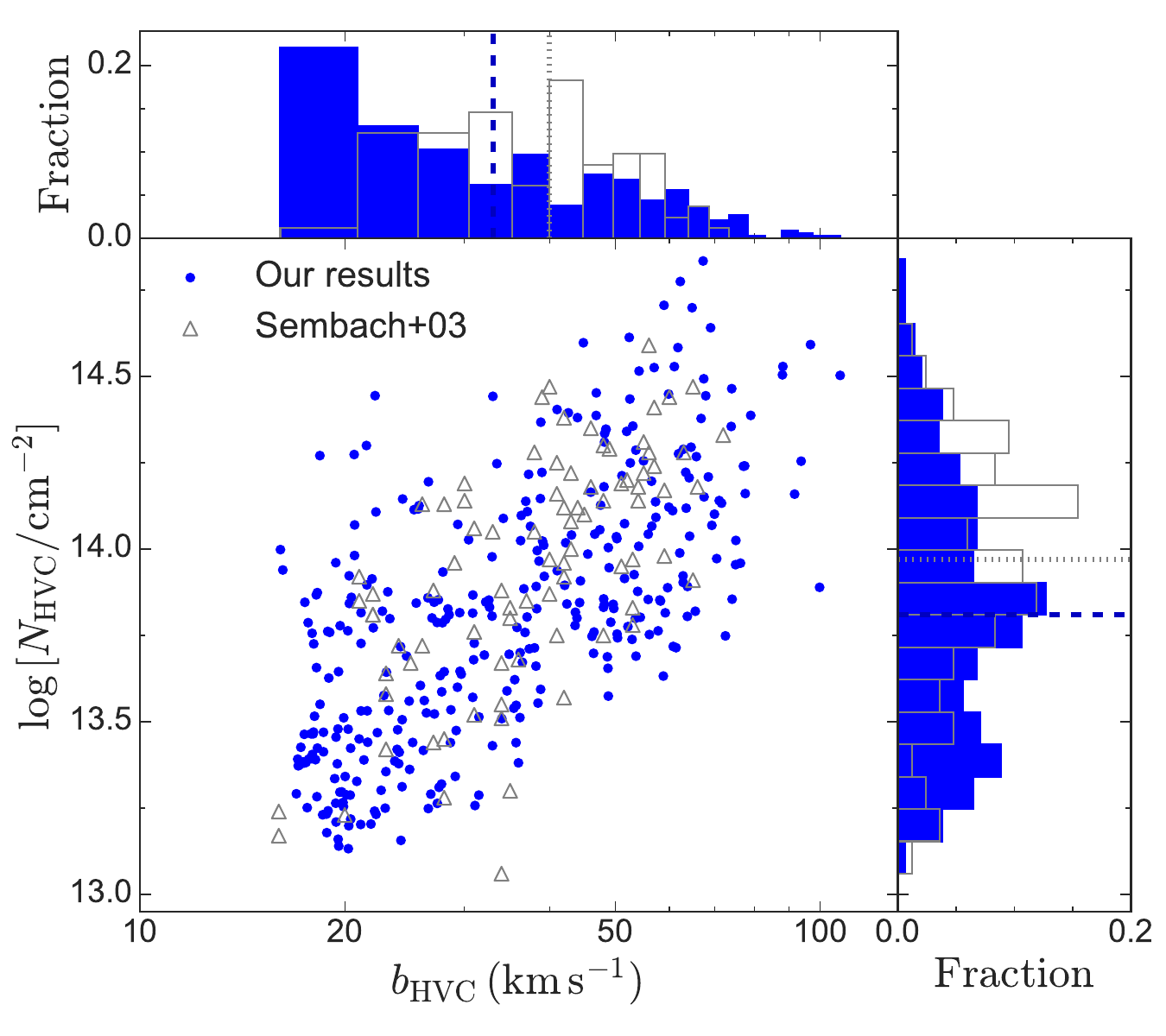}
\caption{Similar to Figure~\ref{fig-Nb-lo} but for high-velocity \ion{O}{6}. The gray triangles denotes {\it FUSE} observations of HVCs \citep{2003ApJS..146..165S}. 
\label{fig-Nb-hi}}
\end{figure}

For HVCs, \ion{O}{6} $1032 \rm \AA$ absorptions have been detected by {\it FUSE} at $\ge 3\sigma$ confidence levels along $59$ out of the $102$ sightlines, among which $100$ are toward extragalactic objects and two toward halo stars \citep{2003ApJS..146..165S}. To make a direct comparison with their results, 
we randomly generate a total of $16\times59=944$ sightlines from four off-center observers, where $59$ is the number of sightlines with detected \ion{O}{6} absorption reported by \citet{2003ApJS..146..165S}.  The background quasars are placed at a distance of $260\, \rm kpc$, and the galactic latitude is limited to $|b|>20^\circ$. 
Accounting for the detected high-velocity \ion{O}{6} properties \citep[see Table 1 in][]{2003ApJS..146..165S}, our mock detections need to satisfy the following conditions: (i) the integration interval $v_{+} -  v_{-} \ge 50$\,km~s$^{-1}$; (ii) \ion{O}{6} column density $\log (N_{\rm HVC}/\rm cm^{-2}) \ge 13.06$; and (iii) \ion{O}{6} line width $b_{\rm HVC} \ge 16\, \rm km~s^{-1}$. This results in 339 detections (the blue filled circles in Figure~\ref{fig-Nb-hi}) of high-velocity \ion{O}{6} out of the $944$ sightlines, with a detection rate (339/944) lower than that (84/102) given by observations.

Our column densities and line widths of high-velocity \ion{O}{6} occupy similar parameter space as the observations \citep{2003ApJS..146..165S}, with $b\sim16-107\ \rm km~s^{-1}$ and $\log (N/\rm cm^{-2})\sim 13.1 -14.8$. The median values are also consistent with the real data considering the $1\sigma$ uncertainties, i.e., $b \sim 33.0\pm18.8$ vs. $40.0\pm13.1\ \rm km~s^{-1}$, and $\log (N/\rm cm^{-2}) \sim 13.8\pm0.4$ vs. $14.0\pm0.3$ (see the seventh and eighth columns in Table~\ref{tab-3}). Unlike the symmetric distribution for LVCs, the line widths for high-velocity \ion{O}{6} peaks at $b\lesssim20\ \rm km~s^{-1}$, suggesting non-thermal motions for high-velocity \ion{O}{6} might be less significant than its low-velocity counterparts. 
In addition, unlike the random distribution for LVCs, there is a significant positive correlation between the column density and Doppler $b$-value for high-velocity \ion{O}{6}, still in line with the {\it FUSE} observations of the MW \citep{2003ApJS..146..165S}. 
Such a correlation may support collisional ionization instead of photoionization as the dominant mechanism for the production of high-velocity \ion{O}{6} \citep[see e.g.,][]{2002ApJ...577..691H}.   
Moreover, photoionization models underproduce observed OVI column densities by order of magnitude \citep[e.g.,][]{2003ApJS..146..165S}, also landing support to the collisional ionization origin.

\subsection{Cumulative column density}
\label{sec:cumulative}

We note that \citet{2020ApJ...896..143Z} investigated cumulative \ion{O}{6} column densities from an inside-out view of MW analogs selected from the Figuring Out Gas \& Galaxies In Enzo (FOGGIE) simulation \citep{2019ApJ...873..129P}. To make a direct comparison with their results, we adopt the same method as that of \citet{2020ApJ...896..143Z} and randomly generate a total of 1000 LOS with $|b| > 20^\circ$ for the four off-center observers. For each of the sightline, we calculate the column density as a function of the distance $r$ to the observer by integrating Equation~(\ref{eq-O6}) over $r$.

\begin{figure}
\centering
\includegraphics[width=0.48\textwidth]{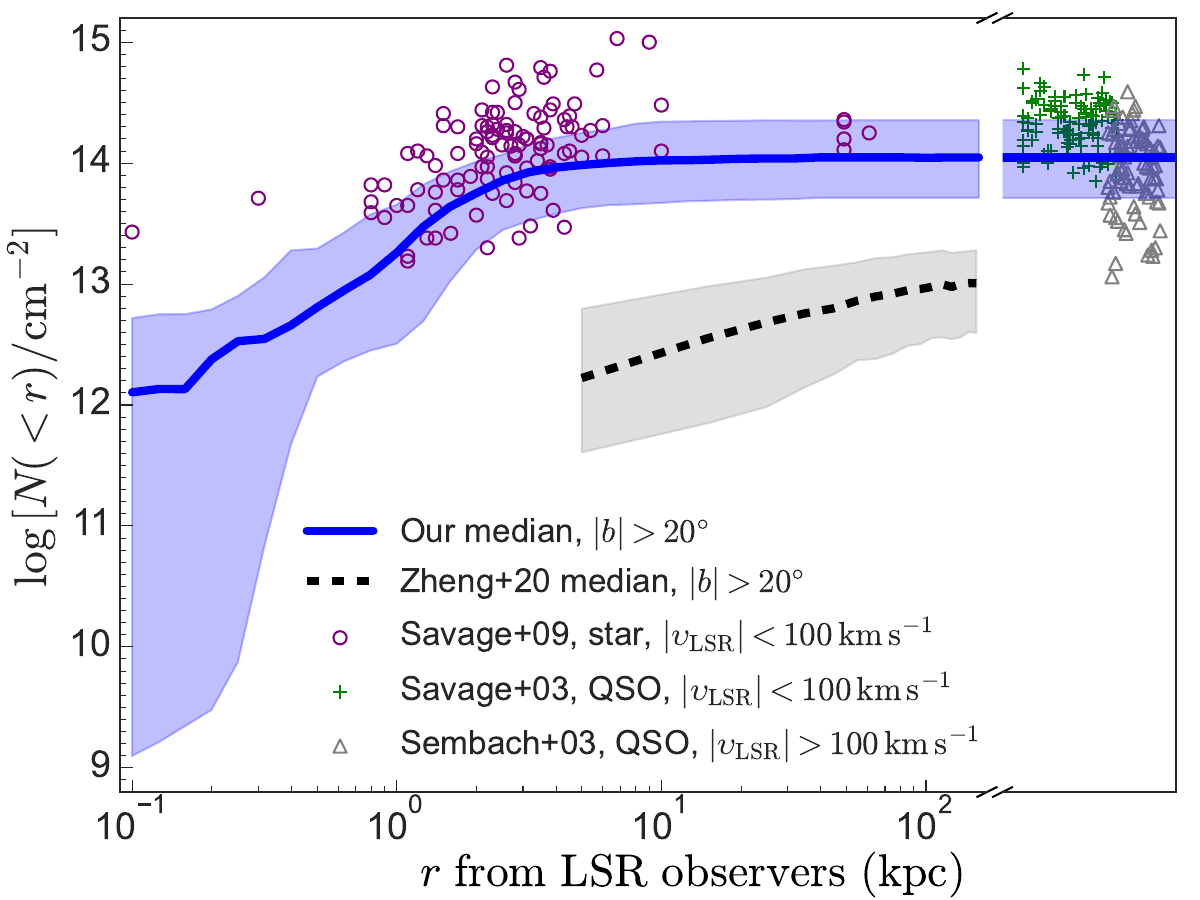}
\caption{Profile of the cumulative \ion{O}{6} column density for the SFE1 run, as represented by the blue solid line and the blue band. Purple open circles,  green pluses, and gray triangles are observations of low-velocity \ion{O}{6} toward stars \citep{2009ApJ...702.1472S}, toward quasars \citep{2003ApJS..146..125S}, and high-velocity \ion{O}{6} toward quasars \citep{2003ApJS..146..165S}.  The gray dashed line and gray band are the simulation results  given by \citet{2020ApJ...896..143Z}. 
\label{fig-Nr}}
\end{figure}

The median profile as well as the $16{\rm th}$ and $84{\rm th}$ percentiles are displayed as the blue solid line and band in Figure \ref{fig-Nr}. 
Despite a systematic offset between our results (blue solid line) and observations of LVCs toward quasars/stars \citep[green crosses and magenta circles;][]{2003ApJS..146..125S, 2009ApJ...702.1472S}, about half of the observational data points are consistent with our 1-sigma uncertainties (blue band). 
The extrapolation of our results to larger distances, i.e., $\log (N/\rm cm^{-2}) \sim 14$, also agrees with HVC observations toward quasars/stars \citep{, 2003ApJS..146..165S} at $r>100\ \rm kpc$. Given that those observations only include low- or high-velocity \ion{O}{6}, each set of the observations may represent a lower limit when compared to our results. The large discrepancy at smaller distances ($r\lesssim0.3\ \rm kpc$) could arise from small-scale clumps and cavities in the ISM induced by SN explosions and other feedback processes \citep{2020MNRAS.499.5862L}, despite the small number statistics.

In contrast, \citet{2020ApJ...896..143Z} underproduced \ion{O}{6} in the halos by $1-2$ orders of magnitude in the column density (the gray dashed line in Figure~\ref{fig-Nr}), comparing to our results and to the observations.
The reason, as they have pointed out, could be that their simulated dark matter halos are smaller than the real case \citep{2016ARA&A..54..529B}, and/or that they only consider the thermal feedback from SNe, which is unable to expel enough metals into the ISM/CGM. The consideration of the full suite of feedback processes (e.g., stellar winds, radiative feedback, and SN explosions) by the {\it SMUGGLE} model and by the fiducial run of L20's simulation could be responsible for our agreement with the real data.

 \subsection{Other simulation models}  
 \label{sec:other}

Results presented in Section~\ref{sec:lvc} to \ref{sec:cumulative} are derived from the fiducial SFE1 run of L20's simulation. To explore how O VI absorption features are affected by different subgrid models, we consider the other five variations listed in Table~\ref{tab-2} for comparison. 

\begin{figure*}
\centering
\includegraphics[width=0.98\textwidth]{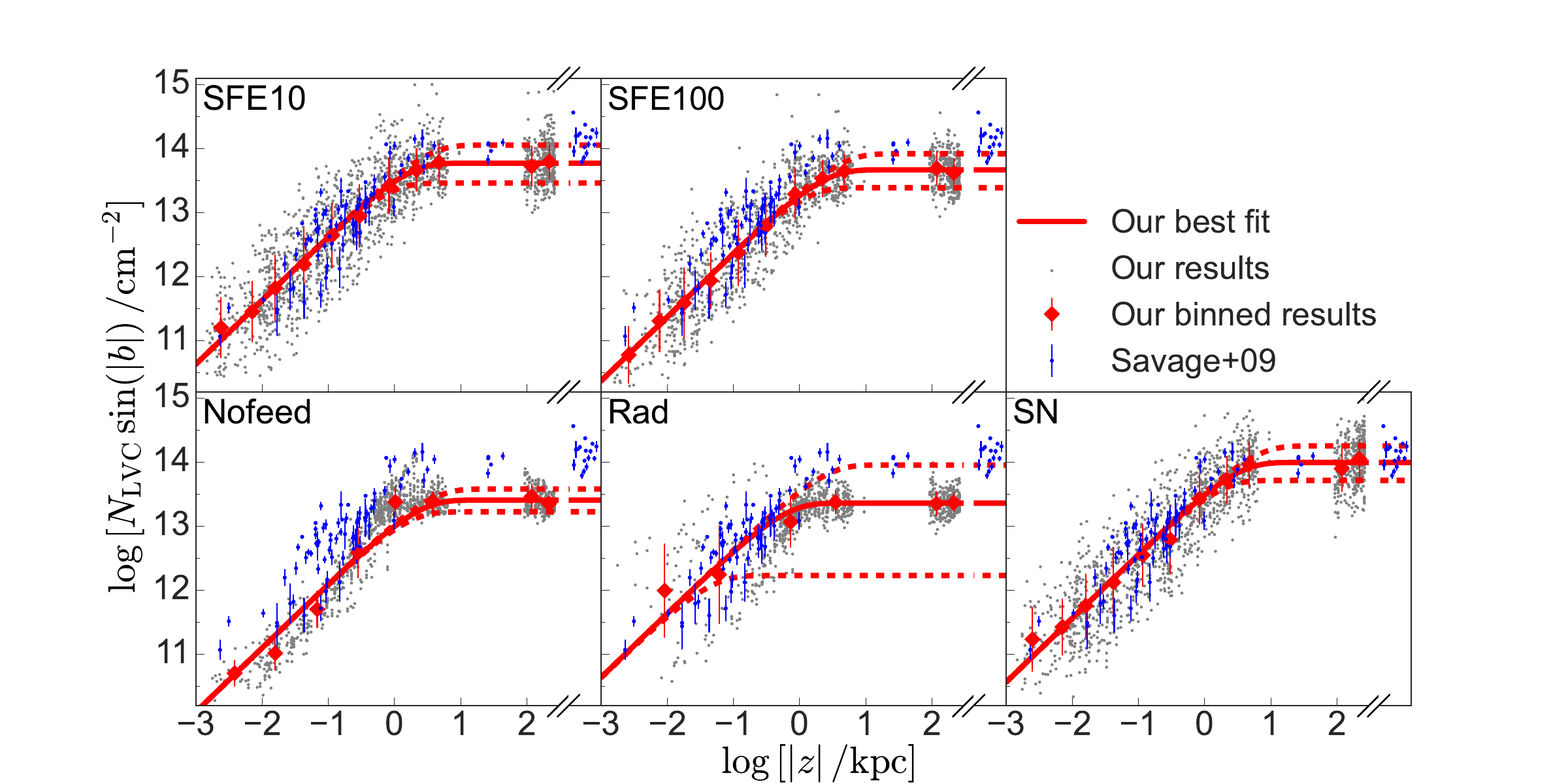}
\caption{Distribution of the projected column densities along $z$-axis vs. heights above the galactic disk for low-velocity \ion{O}{6}, for the other five runs as labeled in each panel.  Legends are similar to those in Figure~\ref{fig-Nz}.
\label{fig-Nz5}}
\end{figure*}

Similar to Figure~\ref{fig-Nz} for the SFE1 run, Figure~\ref{fig-Nz5} shows $\log(N_{\rm LVC} \sin|b|)$ versus $\log|z|$ for the other five models.  While the scale height for low-velocity \ion{O}{6} derived from the fiducial SFE1 run is comparable to the observations \citep{2009ApJ...702.1472S}, the runs with higher star formation efficiency, e.g., SFE10 and SFE100 runs, result in smaller scale height for \ion{O}{6}-bearing gas. 
The scale height does not always decrease with increasing star formation efficiency, which is attributed to the degeneracy between the scale height ($h$) and mid-plane density ($n_0$).  Meanwhile, the projected column density $\log(n_0 h)$ at $|z|\gg h$ decreases slightly as the star formation activity weakens. The reason is that early feedback (e.g., stellar winds, radiation pressure) that is enhanced by intense star formation blows gas and metals away. Higher star formation efficiency also leads to more SN events at a given time-step, and SN feedback could also play a role.   
The Rad run considering only radiative feedback and stellar winds results in much lower scale height and projected column density, compared to the SFE1 run with the full suite of stellar feedback, revealing that SN feedback plays an important role in reproducing the observed spatial distribution of low-velocity \ion{O}{6}, i.e., SN energy and momentum injections collisionally ionize more \ion{O}{6} and push the warm gas further out of the galactic disk. Indeed, the fitting result for the SN run is in nice agreement with the observations. In contrast, the run without feedback $-$ ``Nofeed'', gives an overall lower density and a low scale height for low-velocity \ion{O}{6}.

\begin{figure*}
\centering
\includegraphics[width=0.55\textwidth]{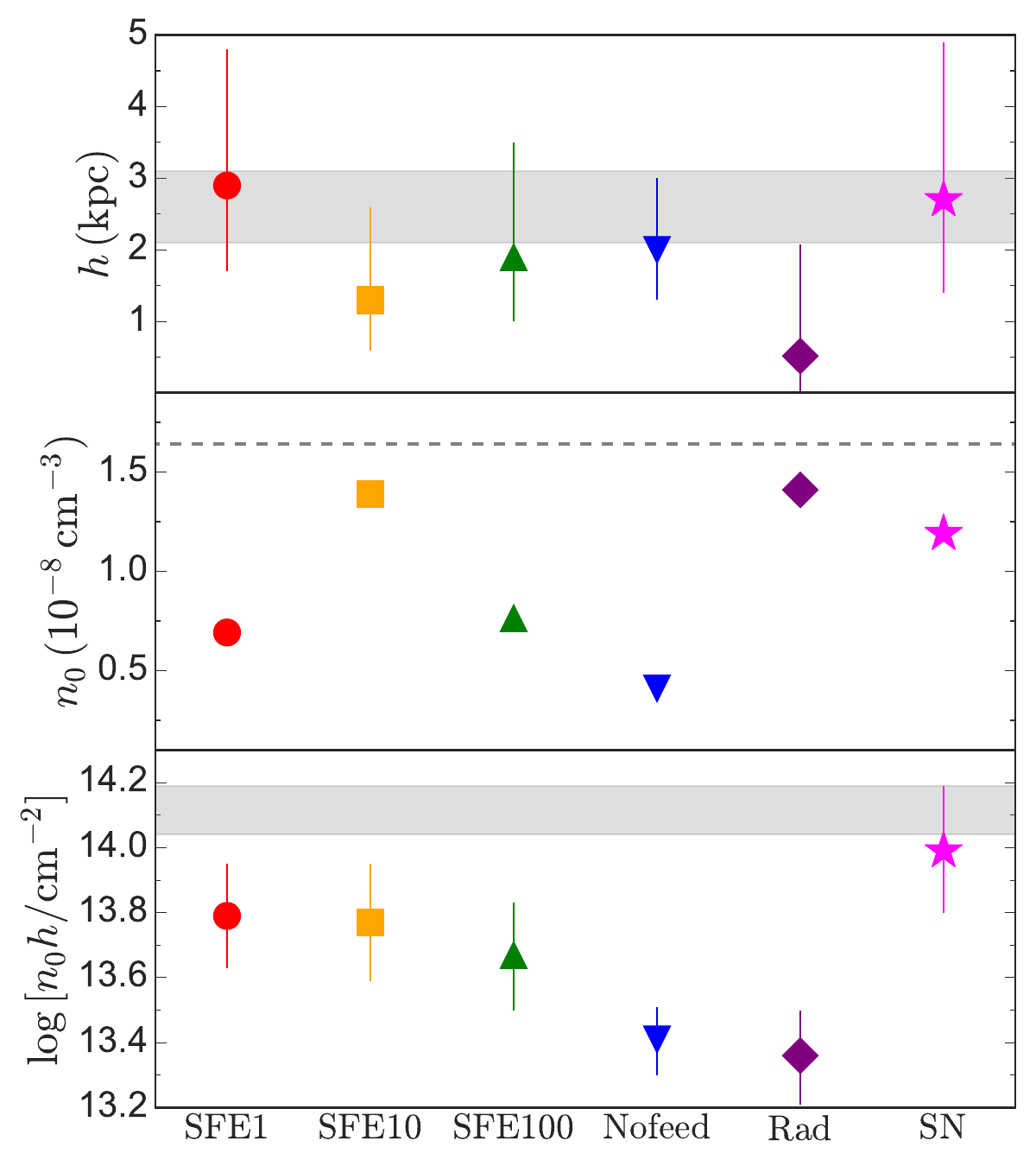}
\caption{Comparison of the best-fit parameters ($y-$axes of the panels) of the disk model for the six model variations ($x-$axis). The gray band and dashed line are  the observational constraints provided by \citet{2009ApJ...702.1472S}. See also the red lines in Figure~\ref{fig-Nz} (the SFE1 run) and Figure~\ref{fig-Nz5} (the other five runs) for the fitting results and Table~\ref{tab-3} for the best-fit parameters.  
\label{fig-model}}
\end{figure*}

Table~\ref{tab-3} (the second to fourth columns) lists the best-fit parameters for the six runs and the values derived from observations \citep{2009ApJ...702.1472S}, which, for a more clear view, are compared in Figure~\ref{fig-model}. 

\begin{figure*}
\centering
\includegraphics[width=0.9\textwidth]{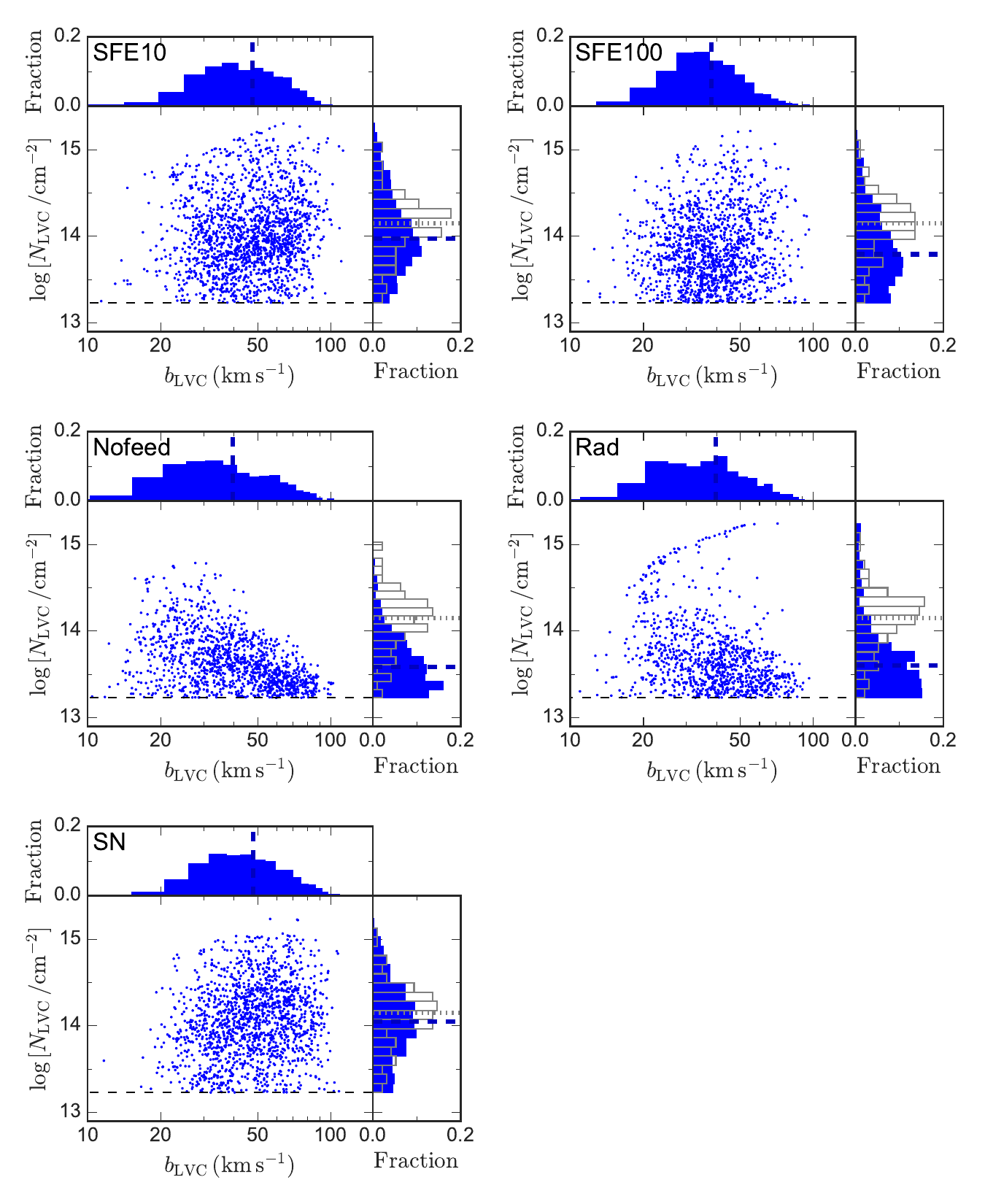}
\caption{The column density and line width distribution for low-velocity \ion{O}{6}, for the other five runs as labeled on the top left of each panel. Legends are similar to Figure~\ref{fig-Nb-lo}.  
\label{fig-Nb5-lo}}
\end{figure*}

Figure \ref{fig-Nb5-lo} displays column densities and line widths distribution of low-velocity \ion{O}{6} for the other five model variations, which are similarly obtained as that for the SFE1 run shown in Figure~\ref{fig-Nb-lo}.  
Similar to the SFE1 run, none of the five runs exhibit obvious correlations. The distribution and median value of \ion{O}{6} column density for the SN run agree excellent well with the observations.  In contrast, the runs lack of SN feedback (e.g., Nofeed and Rad runs) underproduce \ion{O}{6}, with median column densities $\sim0.6$ dex lower. The critical impact of SN feedback is once again highlighted. 

\begin{figure*}
\centering
\includegraphics[width=0.9\textwidth]{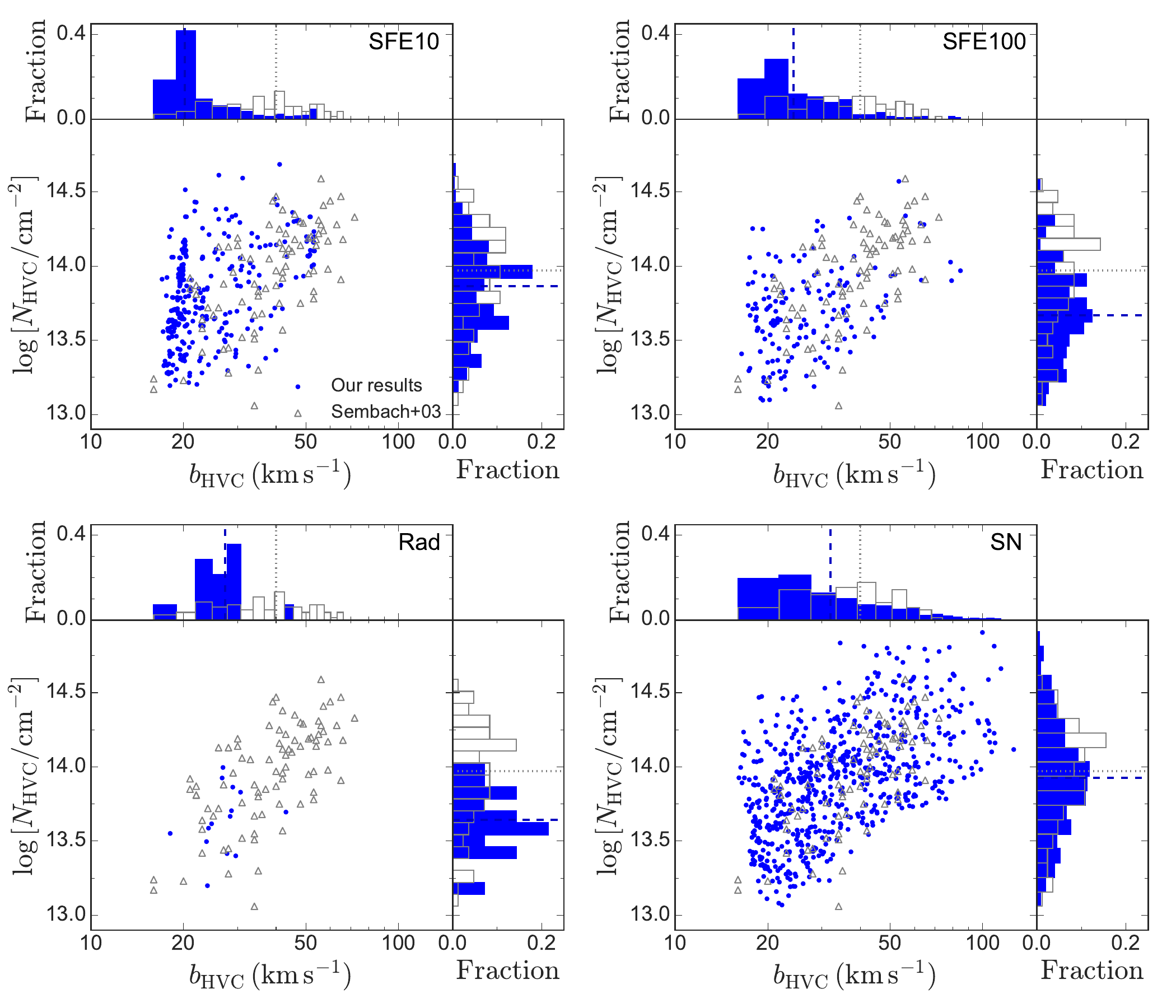}
\caption{Column densities vs. line width distribution for high-velocity \ion{O}{6} for the other four model variations. Legends are similar to Figure \ref{fig-Nb-hi}. The Nofeed run is not displayed because no high-velocity \ion{O}{6} absorptions are detected. 
\label{fig-Nb5-hi}}
\end{figure*}

For high-velocity \ion{O}{6}, the column density $-$ line width relation are displayed in Figure~\ref{fig-Nb5-hi} for the other four runs, as compared to that of the SFE1 run shown in Figure~\ref{fig-Nb-hi}. The median values of the column densities and line widths for different runs are listed in the seventh and eighth columns of Table~\ref{tab-3}. The Nofeed run is not displayed because no high-velocity \ion{O}{6} components are detected, indicating the necessity of feedback processes to accelerate \ion{O}{6} particles. While the column densities of high-velocity \ion{O}{6} derived from different runs generally agree with the observations \citep{2003ApJS..146..165S} accounting for the uncertainties, the median values and distributions of the Doppler parameter support the SFE1 and SN runs, both including SN feedback.  

\begin{figure*}
\centering
\includegraphics[width=0.8\textwidth]{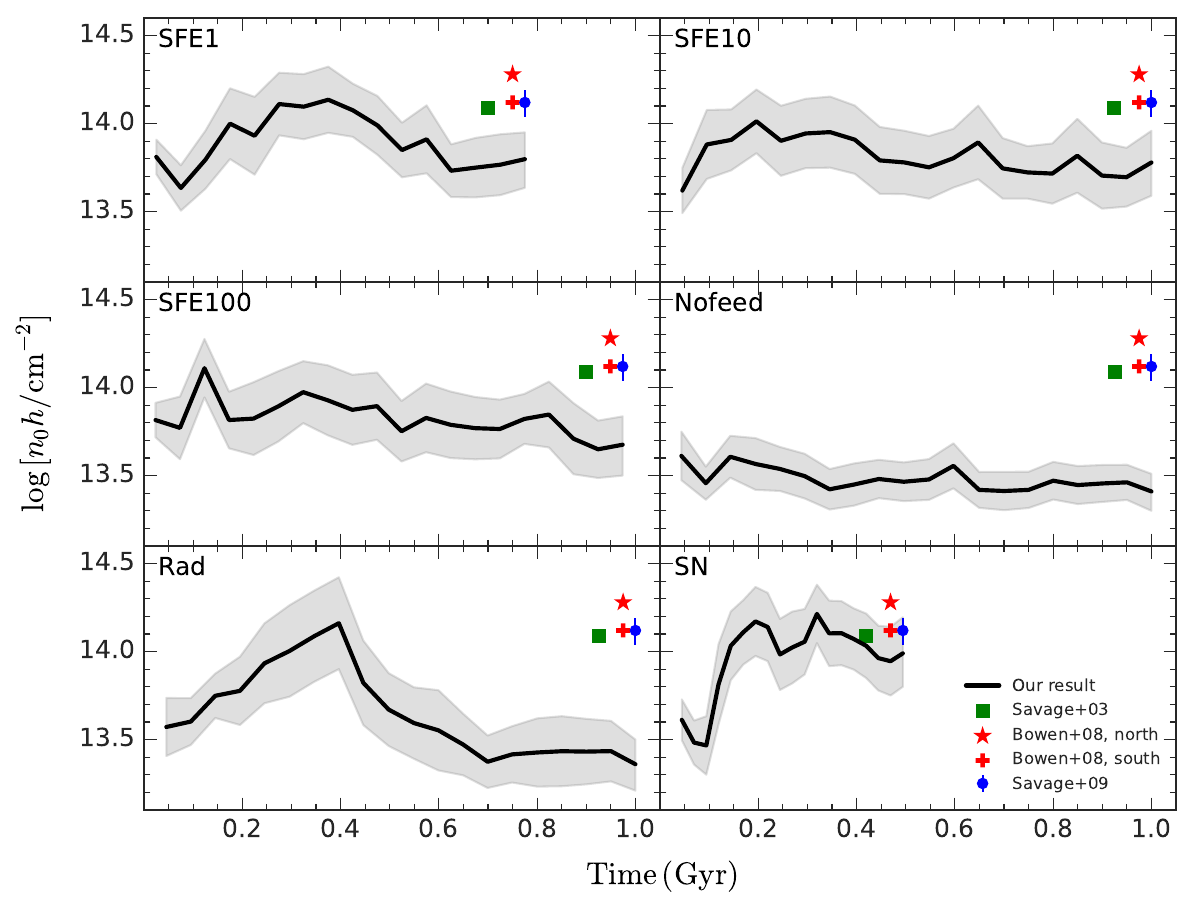}
\caption{The evolution of $\log (n_0 h)$ as a function of time for the six model variations. The black solid line is our results, with the 1$\sigma$ uncertainty represented by the gray region. The symbols denote the observational $\log (n_0 h)$ values of the Galaxy reported by \citet{2003ApJS..146..125S}, \citet{ 2008ApJS..176...59B}, and \citet{2009ApJ...702.1472S}, as labeled, which is arbitrarily shifted along the $x$-axis to make the comparison more clearly.
\label{fig-Nt}}
\end{figure*}

The $\log(n_0 h)$ values for different model variations listed in Table~\ref{tab-3} represent the simulated galaxy at the ``present" time when the simulation is terminated. 
In fact, for each of the snapshot of the simulation, we can similarly obtain its $\log(n_0 h)$ value. In Figure~\ref{fig-Nt}, we present the evolution of $\log(n_0 h)$ across the  simulation time for the six runs. As can be seen by comparing the SFE1, SFE10, and SFE100 runs,  a larger star formation efficiency results in a downward tendency of $\log(n_0 h)$ over time. 
This could be attributed partly to the fast conversion of cold gas to stars and thus less oxygen is left for \ion{O}{6} production via heating. Meanwhile, the stellar winds from young massive stars have an important impact on the ISM gas \citep{1999isw..book.....L, 2012A&A...537A..37M}, e.g, dispersing the gas and impeding the generation of \ion{O}{6} ions via SN feedback heating. 
Consequently, the ``present" value of $\log(n_0 h)$ and its $2\sigma$ confidence region for the SFE1 run marginally agrees with the observations \citep{2003ApJS..146..125S, 2008ApJS..176...59B, 2009ApJ...702.1472S}, yet the SFE100 run deviates further. 
For the SFE1 and SN runs, the simulation data are available only for runtime within $0.8$ and $0.5\, \rm Gyr$, respectively. Based on the currently available data, the ``present" $\log(n_0 h)$ value for these two runs are in better agreement of with the observations than the other four model variations. 

To summarize, comparison of different runs in L20's simulation with the observations of low- and high-velocity \ion{O}{6} favors the SFE1 and SN runs, suggesting that SN feedback is required to reproduce the \ion{O}{6} observations, and meanwhile early feedback associated with star formation activities should be moderate (not too strong), e.g., with star formation efficiency of $\epsilon_{\rm{ff}}\sim0.01$.

\subsection{Comparison with external galaxies} 
\label{sec:external}

Results in Section~\ref{sec:lvc} - \ref{sec:other} are viewed from off-center observers inside the simulated galaxy. Here we present results for the SFE1 run viewed from an external observer and compare with observations of external galaxies. 

\begin{figure*}
\centering
\includegraphics[width=0.9\textwidth]{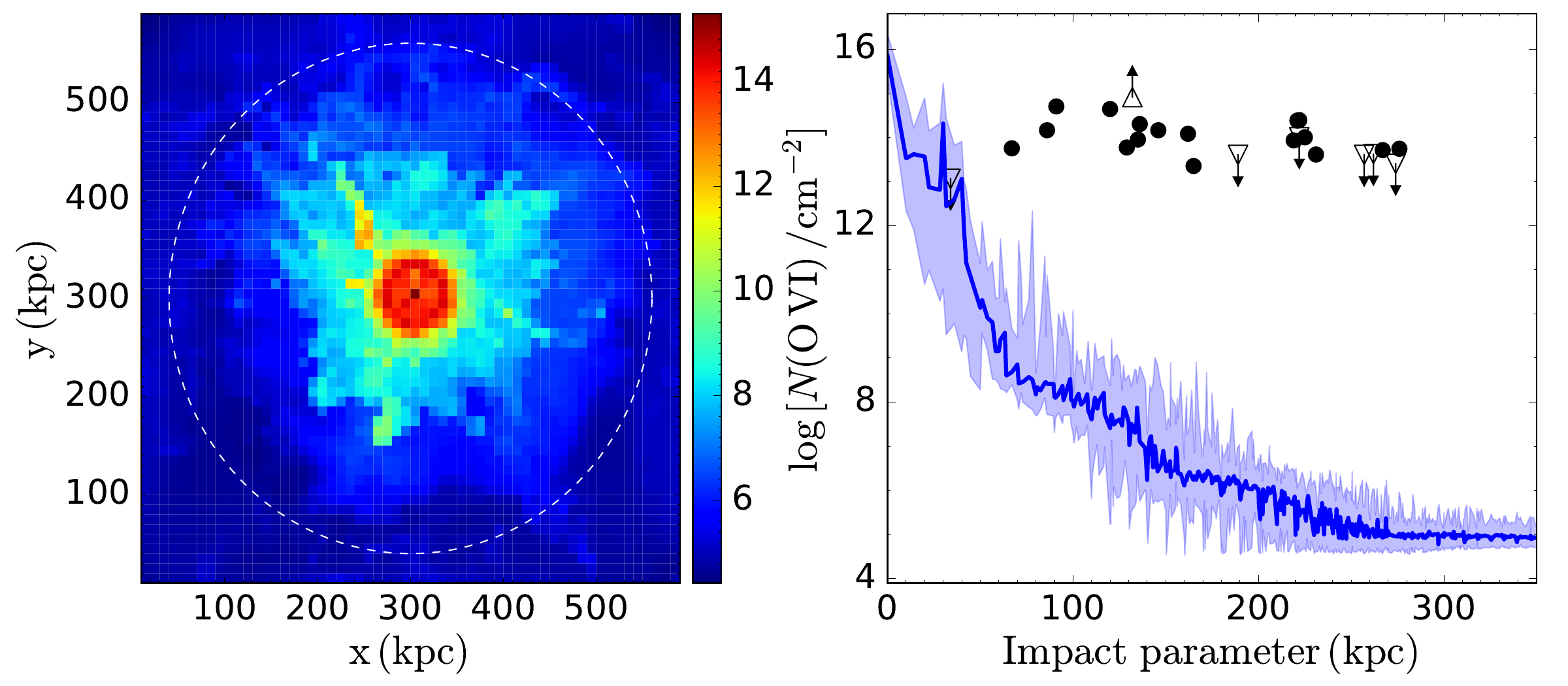}
\caption{{\it Left:} Face-on projection of \ion{O}{6} column density (along $z$-axis) for the SFE1 run, as the color bar denotes. The white dashed line marks the galaxy's virial radius of $260\, \rm kpc$. {\it Right:} \ion{O}{6} column density as a function of the impact parameter derived from random face-on and edge-on sightlines. The blue solid line is our median result, and the blue area denotes the $5{\rm th}-95{\rm th}$ percentiles range. The black filled circles and open triangles are the observational results for sub- and super-$L_*$ galaxies given by \citet{2011ApJ...740...91P}.  
\label{fig-sden}}
\end{figure*}

The left panel of Figure~\ref{fig-sden} shows the face-on view of \ion{O}{6} column density map (on $xy-$plane). For each grid of coordinates ($x,\ y$), the column density is obtained by integrating \ion{O}{6}  number density in Equation~(\ref{eq-O6}) along the $z$-axis with path length of $600\ \rm kpc$, i.e., the size of the simulation box. The white dashed line denotes the virial radius of $260\, \rm kpc$. The column density peaks at the center with $\log (N/\rm cm^{-2}) \sim 15.3$ and gradually declines toward outer region, approaching a background value of $\log (N/\rm cm^{-2}) \sim 4.4$. Besides that, there is tentative evidence for structures spanning tens of $\rm kpc$. 

To make a direct comparison with observations of external galaxies, we plot the column density versus the impact parameter in the right panel of Figure~\ref{fig-sden}. To achieve that, we generate random sightlines for both face-on and edge-on views of the simulated galaxy. The blue solid line shows the median column density, and the blue band shows the range of $5{\rm th}$ to $95{\rm th}$ percentiles. Our results are consistent with the observations of sub- and super-$L_*$ galaxies \citep{2011ApJ...740...91P} for impact parameter $\lesssim50\, \rm kpc$. Beyond that, the column density sharply declines and drops below the observational values. This happens as expected because  the simulation performed by L20 as well as the {\it SMUGGLE} galaxy formation model is for an isolated galaxy without gas supply from the IGM, which is also the shortcoming of this study. Indeed, the Galactic halo density ($\sim10^{-4}\, \rm cm^{-3}$)  suggested by observations of the Magellanic Stream  \citep{1996AJ....111.1156W} is more than one order of magnitude higher than our results ($\lesssim10^{-5}\, \rm cm^{-3}$; the panel (c) of Figure~\ref{fig-Tvn}) at a radius of $\sim 50\, \rm kpc$ .

\subsection{Caveats} 
\label{sec:caveat}

\subsubsection{Isolated galaxy simulation}
\label{sec:isolate}

Our results are based on L20's simulations for an isolated galaxy without gas fueling from the IGM and interactions with companion galaxies. This could lead to an underestimation of \ion{O}{6} column density at outer regions, e.g., $r\gtrsim 100\, \rm kpc$ (see Figure~\ref{fig-sden}). In addition, our high-velocity \ion{O}{6} clouds can only be produced via the galactic fountain, i.e., triggered by stellar feedback \citep[e.g.,][]{1976ApJ...205..762S, 1980ApJ...236..577B, 2006MNRAS.366..449F}.  If other mechanisms such as accretion from the IGM \citep[e.g.,][]{2015MNRAS.447L..70F, 2009ApJ...700L...1K} and materials stripped or ejected from satellites \citep[e.g.,][]{2004ASSL..312..101P, 2013A&A...550A..87H}, are also responsible for the formation of high-velocity \ion{O}{6}, our simulation (Table~\ref{tab-3}) may underproduce high-velocity \ion{O}{6} content and distort its spatial distribution.

\subsubsection{The metallicity}

\begin{figure}
\centering
\includegraphics[width=0.48\textwidth]{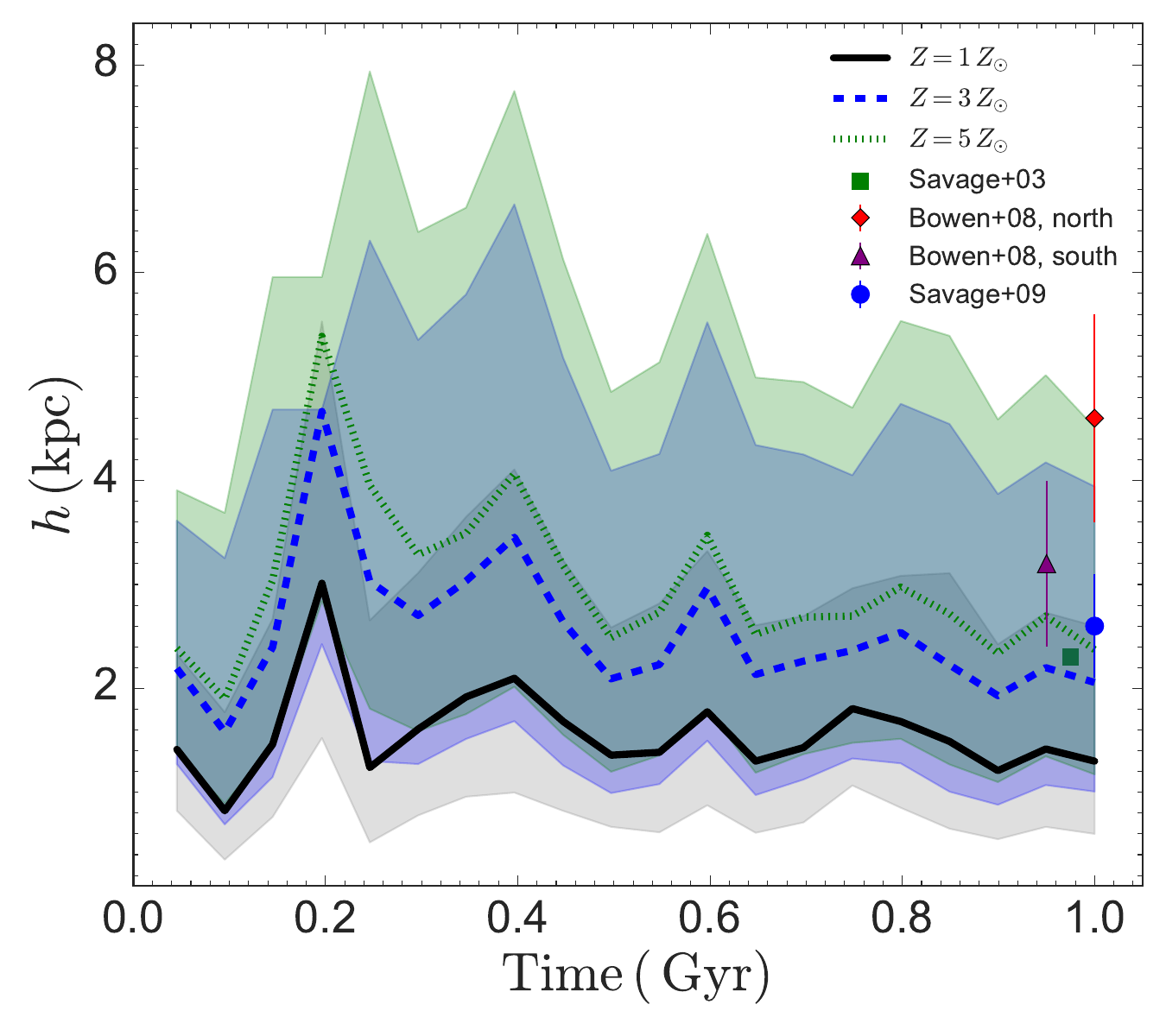}
\caption{The evolution of the exponential scale height $h$ of low-velocity \ion{O}{6} as a function of time for the SFE10 run with metallicities of $1\, Z_\odot$ (black solid line), $3\, Z_\odot$ (blue dashed line), and $5\, Z_\odot$ (green dotted line), respectively. The lines represent the median values for random sightlines described in Section~\ref{sec:lvc}, and the regions with corresponding colors represent $1\sigma$ uncertainties. The symbols with error bars are observational results reported by \citet{2003ApJS..146..125S}, \citet{ 2008ApJS..176...59B}, and \citet{2009ApJ...702.1472S}, as labeled. In particular, \citet{2008ApJS..176...59B} provides the scale heights of low-velocity \ion{O}{6} for the northern ($b>20^\circ$) and southern ($b<-20^\circ$) hemisphere of the MW, respectively. 
\label{fig-mat}}
\end{figure}

Our results in this work are obtained under the assumption of solar metallicity for the gas when converting the number density of hydrogen to that of \ion{O}{6} in Equation~(\ref{eq-O6}). Constant metallicity is often assumed for simplicity despite the fact that the metallicity could differ by orders of magnitudes for different regions of the galaxy \citep[e.g.,][]{2017MNRAS.464.2796G, 2021Natur.597..206D}.  
Alternatively, we quantify the effect of metallicity on the scale height evolution of low-velocity \ion{O}{6} for the SFE10 run in Figure~\ref{fig-mat}, by setting three constant metallicities of $1\, Z_\odot$, $3\, Z_\odot$, and $5\, Z_\odot$. As expected, a higher metallicity results in a larger scale height of \ion{O}{6}, which differs by a factor of $\lesssim 2$ for $1\, Z_\odot$ and $5\, Z_\odot$ cases, comparable to the variations of the scale height across the simulation time of $\sim1\, \rm Gyr$. While the SFE10 run is ruled out under the assumption of solar metallicity when compared to the observations \citep{2003ApJS..146..125S, 2008ApJS..176...59B, 2009ApJ...702.1472S}, higher metallicity of $5\, Z_\odot$ makes the SFE10 run's results ($h=2.3_{-1.2}^{+2.2}\, \rm kpc$) well consistent with the observations considering the errors. This indicates that  to some extent, a higher metallicity can compensate for lower \ion{O}{6} content caused by strong early feedback (e.g, stellar winds, radiation pressure) launched by short-lived massive stars. 

\subsubsection{The UV background and other ionizing sources}

Our results on \ion{O}{6} properties of the simulated galaxies are based on Equation~(\ref{eq-O6}), where the ionization fraction of \ion{O}{6} is derived via {\sc Cloudy} \citep{2017RMxAA..53..385F} modeling by taking into account extragalactic UV background radiation \citep{1996ApJ...461...20H}. While such UV background is typically applied to intergalactic medium regions \citep[e.g.,][]{2001ApJ...561L..31F}, there are alternative versions of the UV background in the literature and other potential contribution of ionizing sources, e.g., stellar radiation within the galaxy and cosmic ray heating \citep{2014ApJ...792....8W}. 

The spectral shape of UV background has been shown to affect oxygen abundance \citep{2008ApJ...689..851A} and statistics of \ion{O}{6} absorbers in the IGM \citep{2009MNRAS.395.1875O}. A comparison of various UV background have been presented in the Figure 1 of \citet{2023MNRAS.523.2296M}, including the one \citep{1996ApJ...461...20H} we adopted. 
Energy of $\sim114\, \rm eV$ required for photoionizing \ion{O}{5} corresponds to the high-energy tail of the spectral energy distribution (SED) of background radiation field. Consequently, most \ion{O}{6} could be produced from collisional ionization at temperatures of $\sim3\times10^5\, \rm K$ rather than from photoionization at lower temperatures \citep[e.g.,][]{2005ARA&A..43..337C}. Moreover,
the flux difference at the ionizing energy is at most $\sim0.5\, \rm dex$ for various frequently used UV background and should not make much difference.

Photoionization is considered as a channel of radiative stellar feedback in the framework of the {\it SMUGGLE} model, and  \ion{O}{6} distribution that is controlled by the ionization fraction $f_{\rm O\ \Romannum{6}} (T, n_{\rm H})$ in  Equation~(\ref{eq-O6}) is affected by feedback processes in terms of heating (increasing the temperature $T$) and/or blowing gas away (decreasing hydrogen density $n_{\rm H}$). However, the stellar radiation inside the galaxy is not directly considered in the {\sc Cloudy} modeling. 
 The contribution of a starburst galaxy to the total ionizing photons is evaluated in the Figure 13 of \citet{2014ApJ...792....8W}. The SED of the radiation field with the contribution of the starburst galaxy  with a star formation rate (SFR) of $1\, \rm M_\odot \ yr^{-1}$ at a distance of $d=72\, \rm kpc$ has similarly flat slope toward higher energies ($E\gtrsim70\, \rm eV$), and deviates $\sim0.1\, \rm dex$ from the \citet{2001cghr.confE..64H} UV background. 
While the assumed SFR is typically true for our simulated MW-like galaxy \citep{2020MNRAS.499.5862L}, the distance of \ion{O}{6} clouds to the star forming region spans a wide range  across the halo ($0-260\, \rm kpc$). Figure 8 of \citet{2005ApJ...630..332F} compares the UV background radiation with the radiation from the MW at different locations within the Galaxy, and reveals that the radiation field from the Galaxy is similar to the \citet{1996ApJ...461...20H} UV background at $E\sim114\, \rm eV$ for a distance of $d\sim20-30\, \rm kpc$.  
The sharp decrease of the radiation flux at $54\, \rm eV$ arising from \ion{He}{2} edge in hot stars indicates that high-energy photons of $E>114\, \rm eV$ are very limited. Detailed {\sc Cloudy} modeling suggests that photoionization is negligible for the production of \ion{O}{6} despite its dependence on the radiation field adopted.

Cosmic ray heating could serve as a crucial supplementary source of ionization and heating within the Galactic virial radius \citep{2013ApJ...767...87W}. For gas densities $\gtrsim10^{-2}$\,cm$^{-3}$, the cosmic ray background (CRB) can dominate  over photoelectric heating for gas accounting for a weaker dependence on the gas density for the CRB heating. Therefore, CRB could significantly enhance the density of \ion{O}{6} in low density regions, although the precise number is challenging to determine because of the poorly constrained local CRB \citep[see discussions in][]{2014ApJ...792....8W}.

\section{Conclusions} 
\label{sec:summary}

We study \ion{O}{6} properties in MW-like galaxies by analyzing the suites of simulation performed by L20 in the framework of {\it SMUGGLE} galaxy formation model. We find that the {\it SMUGGLE} model is capable of producing consistent global properties of Galactic warm gas traced by \ion{O}{6}.
In addition, mechanical stellar feedback is shown to have a crucial impact on the spatial distribution and kinematics of \ion{O}{6} absorbers.  Particularly, SN feedback is necessarily required, and early feedback associated with star formation activities needs to be moderate to reproduce \ion{O}{6} observations. 
Our main findings are detailed as follows.

\begin{enumerate}[label=(\roman*), align=left]
	\item Low-velocity \ion{O}{6} distribution is well described by an exponentially declining disk with a scale height of $2.9^{+1.9}_{-1.2}\, \rm kpc$ and $\log(n_0h/{\rm cm^{-2}}) = 13.79\pm0.16$ for the fiducial SFE1 run (with full suites of feedback processes), generally consistent with the observations. The SN run (with SN feedback only) results in a scale height well consistent with observations as well. Other runs turning off SN feedback or with higher star formation efficiencies lead to smaller values for the scale height.    
	
	\item For the SFE1 run, the column density of low-velocity \ion{O}{6} is distributed in the range of $\log (N/\rm cm^{-2}) \sim 13.2 - 15.2$ with a median value of $\sim13.8$, consistent with observations within $1\sigma$ uncertainties. The line width of low-velocity \ion{O}{6} follows a Gaussian-like distribution over $b\sim 13 - 106\, {\rm km\, s^{-1}}$ with a median value of $47.4\, {\rm km\, s^{-1}}$. No correlations are found between the column density and line width of low-velocity \ion{O}{6} for all of the model variations. 
	
	\item For high-velocity \ion{O}{6} in the SFE1 run, the column density spans $\log (N/\rm cm^{-2}) \sim 13.1 - 14.8$ with a median of $\sim 13.8$, and line width covers $b\sim 16 - 107\, {\rm km\, s^{-1}}$ with a median of $\sim 33\, {\rm km\, s^{-1}}$. A positive correlation are found between the column density and line width of high-velocity \ion{O}{6}, supporting collisional ionization as the dominant mechanism for the production of high-velocity \ion{O}{6}. No high-velocity \ion{O}{6} clouds are found in the run turning off all channels of stellar feedback. 
		
	\item The profile of cumulative \ion{O}{6} column density generally agrees with observations for the SFE1 run.	
The evolution of $\log(n_0h)$ as a function of simulation time also supports the SFE1 and SN runs when comparing to observations. 	
	\item We cannot reproduce observations of column density profile for external galaxies due to the lack of accretion in our simulations, suggesting that accretion is an important part of galaxy evolution modeling.
\end{enumerate}

Overall, the observed Galactic \ion{O}{6} properties can be reasonably reproduced with simulations of isolated MW-like discs based on the {\it SMUGGLE} model with novel treatment of ISM and stellar feedback, in complement to L20's findings of its success in producing realistic cold ISM. A test of its ability in reproducing hotter Galactic gas traced by highly ionized metal species such as \ion{O}{7} and \ion{O}{8} is deferred to a future work.  
One shortcoming of the {\it SMUGGLE} model could be the lack of cosmological gas accretion. The next generation of the {\it SMUGGLE} model intends to involve cosmological simulations with zoom-in of individual objects, and will serve as a powerful tool for predicting galactic structure, outflows, and CGM properties.

\section*{Acknowledgements}
We thank the anonymous referee for his/her helpful comments that have improved the paper.
We are grateful to Greg Bryan for his insightful comments and suggestions.
This work is supported by National Natural Science Foundation of China under grant Nos. 11890692, 12133008, 12221003, 12103017, 12273031, and 11903056, and by the Fundamental Research Fund for the Central Universities (No. 20720230016) of China. We acknowledge the science research grants from the China Manned Space Project, under No. CMS-CSST-2021-A04. 

\software{ 
yt \citep{2011ApJS..192....9T}, 
{\sc Cloudy} \citep{2017RMxAA..53..385F},
Astropy \citep{2018AJ....156..123A}, 
Matplotlib \citep{4160265}, 
SciPy \citep{2020SciPy-NMeth}.
}

\bibliographystyle{aasjournal}
\bibliography{refer}{}
\end{CJK*}
\end{document}